\documentclass{aa}
\usepackage{graphics}

\def\ltapprox{\raise 2pt \hbox {$<$} \kern-1.1em \lower 5pt \hbox {$\approx$}}
\def\ltsim{\raise 2pt \hbox {$<$} \kern-1.1em \lower 4pt \hbox {$\sim$}}
\def\gtsim{\raise 2pt \hbox {$>$} \kern-1.1em \lower 4pt \hbox {$\sim$}}

\begin{document}

\title{{\it Chandra} discovery of extended non-thermal emission in 
3C 207 and the spectrum of the relativistic electrons}


\author{G. Brunetti\inst{1,2}
\and M. Bondi\inst{2}
\and A. Comastri\inst{3}
\and G. Setti\inst{1,2} }
\offprints{G.Brunetti, c/o Istituto di Radioastronomia del CNR, via Gobetti
101, I--40129 Bologna, Italy}

\institute{Dipartimento di Astronomia, 
via Ranzani 1, I--40127 Bologna, Italy
\and
Istituto di Radioastronomia del CNR,
via Gobetti 101, I--40129 Bologna, Italy 
\and
Osservatorio Astronomico di Bologna,
via Ranzani 1, I--40127 Bologna, Italy}

\date{}

\abstract{
We report on the {\it Chandra} discovery of large scale non--thermal
emission features in the double lobed SSRL quasar 3C 207 (z=0.684).
These are: a diffuse emission well correlated with the western radio lobe,
a bright one sided jet whose structure coincides with that of the 
eastern radio jet and an X--ray source at the tip of the jet coincident
with the hot spot of the eastern lobe.
The diffuse X--ray structure is best interpreted as inverse Compton
(IC) scattering of the IR photons from the nuclear source and provides
direct observational support to an earlier conjecture (Brunetti et al., 1997)
that the spectrum of the relativistic electrons in the lobes of
radio galaxies 
extends to much lower energies than those involved in the synchrotron
radio emission.
The X--ray luminous and spatially resolved knot along the jet 
is of particular interest: by combining VLA and {\it Chandra}
data we show that a SSC model is ruled out, while the X--ray spectrum
and flux can be accounted for by the IC scattering of the CMB
photons (EIC) under the assumptions of a relatively strong boosting
and of an energy distribution of the relativistic electrons
as that expected from shock acceleration mechanisms.
The X--ray properties of the hot spot are consistent with a SSC model.
In all cases we find that the inferred magnetic field strength 
are lower, but close to the equipartition values.
The constraints on the energy distribution
of the relativistic electrons, imposed by the X--ray spectra of the
observed features, are discussed.
To this aim we derive in the Appendices precise semi--analytic
formulae for the emissivities due to the SSC and EIC processes.
}

\titlerunning{Extended non--thermal X--ray emission from 3C 207}

\maketitle

\section{Introduction}

A full understanding of the energetics and energy 
distribution of relativistic particles in radio jets and lobes
of radio galaxies and quasars is of basic importance for 
a complete description of the physics and time evolution 
of these sources.
It is assumed that the relativistic
electrons in powerful radio sources are channeled into the 
jet and re--accelerated both in the jet itself
(e.g., in the knots) and in the radio hot spots,  
which mark the location of strong planar shocks formed at
the beam head of a supersonic jet
(e.g. Begelman, Blandford, Rees, 1984). 
Then particles diffuse from the hot spots
forming the radio lobes.

X--ray observations of both compact features and lobes
of radio sources are of basic importance in constraining 
the energetics and spectrum of the relativistic electrons
and possibly their evolution.

Until the advent of {\it Chandra}, only a few examples
of X--ray emission from radio jets and
hot spots were available (e.g. Cygnus A: Harris, Carilli, 
Perley, 1994; 3C 120: Harris et al., 1999;
3C 273: R\"{o}ser, et al., 2000; M87: Neumann, et al., 1997).
It has been immediately clear that the detected emission 
is of non--thermal
origin with the best interpretation provided by  
synchrotron and SSC mechanisms 
under minimum energy conditions
(equipartition between electron and magnetic field energy 
densities).

X--ray emission from the radio lobes has been discovered by ROSAT and
ASCA in a few nearby radio galaxies, namely Fornax A (Feigelson et al.1995;
Kaneda et al.1995; Tashiro et al.2001)
and Cen B (Tashiro et al., 1998),
with the best interpretation provided by the IC scattering of 
the Cosmic Microwave Background (CMB) photons.
In addition combined ROSAT PSPC, ASCA and ROSAT HRI observations
of the powerful radio galaxy 3C 219 have also shown the presence 
of extended X--ray emission within $\sim 80$ kpc distance from the 
nucleus which is interpreted as IC scattering of the IR photons
from a hidden quasar (Brunetti et al.1999).
In general these observations are complicated due to the weak
X--ray brightness and relatively low count statistics, however,
the magnetic field strengths estimated from the combined IC and synchrotron
radio fluxes are within a factor of $\sim 3$ consistent with the
equipartition values. 
Up to now no clear evidence of extended non--thermal emission
from radio loud quasars has been found.

Likewise {\it Chandra} is providing a
significant progress on the study of the X--ray emission 
from jets and hot spots (Chartas et al., 2000;
Harris et al., 2001; 
Wilson, et al., 2000; Schwartz et al., 2000; 
Wilson, et al., 2001;
Hardcastle, et al., 2001a,b;
Pesce et al.2001).
Although the analysis of the 
increasing number of successful detections of
X--ray emission from compact hot spots and 
jets has confirmed the non--thermal nature of the X--rays 
from these sources, it is not clear whether the SSC and
synchrotron model 
can provide or not a general interpretation of
the data (e.g. Harris 2000).
One of the most striking cases is the X--ray jet of
PKS 0637$-$752, where the SSC mechanism can only provide a
flux about 2 orders of magnitude below the observed 
(Chartas et al., 2000). A viable explanation is based on 
the assumption that even far from the core the jet has a
relatively high relativistic bulk motion 
(Celotti, et al., 2001; Tavecchio et al., 2000) 
so that the CMB photons can be efficiently 
up--scattered into the X--ray band (EIC model).

{\it Chandra} is the first X--ray observatory 
able to well resolve the radio lobes of powerful and relatively
distant radio galaxies and quasars; thus, 
for the first time it is now possible
to well separate the non--thermal lobe emission in these objects
from the surrounding cluster emission.
An example is provided by the radio galaxy 3C 295 where the diffuse
X--ray emission from the radio lobes in excess to that of 
the surrounding cluster has been best interpreted
via IC scattering of nuclear photons under equipartition conditions
(Brunetti et al., 2001).

In this paper we report the {\it Chandra} observation of the radio
loud quasar 3C 207.
A powerful one sided X--ray jet, coincident with the radio one is
discovered whereas, in the opposite direction, extended X--ray emission 
coincident with the western radio lobe is also detected.
The X--ray jet is characterized by two knots which coincide
with the main radio knot and with the western hot spot.
In Sects.2.2 and 2.3 we report the radio VLA and {\it Chandra}
results, respectively.
In Sect.3.2 we report the interpretation and modeling
of the results obtained for the lobe emission, while the
interpretation of the western
knot and hot spot is reported in Sects.3.3 and 3.4, respectively.
To model these results we make use of the
electron energy distributions
obtained in the framework of shock acceleration theory and of 
general equations for SSC and EIC processes.
A simple
analytic model for electron 
shock acceleration is worked out in Sect.3.1, whereas 
the general SSC and EIC equations are derived in the Appendices.
The discussion of the results and conclusions are reported in 
Sect.4.

\noindent
$H_0 =50$ km s$^{-1}$ Mpc$^{-1}$ and $q_0 =0.5$ are
assumed throughout: 1 arcsec corresponds to 7.9 kpc at the
redshift of 3C 207.

\section{Target and data analysis}

\subsection{3C 207}

The powerful radio source 3C 207 is identified with 
a quasar at a redshift of z=0.684.
The integrated radio spectrum is relatively steep, 
$\alpha \simeq 0.9$ ($F(\nu) \propto \nu^{-\alpha}$),  
between 150--750 MHz (Laing et al., 1983), 
whereas it flattens at higher frequencies
($\alpha \simeq 0.64$ and 0.43 around 2.5 and 40 GHz, respectively;
Herbig \& Readhead 1992) where the nuclear component dominates.

Bogers et al.(1994, and ref. therein) obtained detailed radio images at
1.4 and 8.4 GHz.
The radio source is relatively compact with an  
extension of about 10 arcsec.
At low radio frequencies a relatively high brightness double lobed 
structure appears; the radio flux and extension of the two lobes 
are similar. 
At high frequency 3C 207 shows a prominent 
radio jet pointing in the direction of the eastern lobe.

The combination of powerful radio lobes and relatively
small angular extension makes this source a good 
candidate to study the IC scattering of the nuclear photons in 
the radio lobes.
In addition, the high spatial resolution of {\it Chandra} allows to 
well resolve the luminous radio jet in the X--ray band and to 
obtain spectral information on both the X--ray jet and lobes.

\subsection{The archive radio Data}

With the aim to compare the radio and
X-ray morphology on the arcsec-scale, and derive the radio spectral 
indices of the extended emission and compact components, we 
requested and re-analyzed some of the VLA archive data (Tab.1)
of 3C 207.
\begin{table}
\caption{VLA Data Archive}
\begin{tabular}{cccccc}
\hline
Proj Code & Date       &$\nu$(MHz) & Array& TOS(s)\\
AL280        & Dec-13-1992& 1415/1665 &  A   & ~3490 \\
AL280        & Dec-13-1992& 14915/14965& A   & ~~540 \\
AL113        & Jul-12-1986& 4835/4885 &  B   & ~~670 \\
AB796        & Nov-08-1996& 8435/8475 &  A   & 16410 \\
AB796        & Mar-04-1997& 8435/8485 &  B   & ~1490 \\
\hline
\end{tabular}
\end{table}

\begin{figure*}
\resizebox{\hsize}{!}{\includegraphics{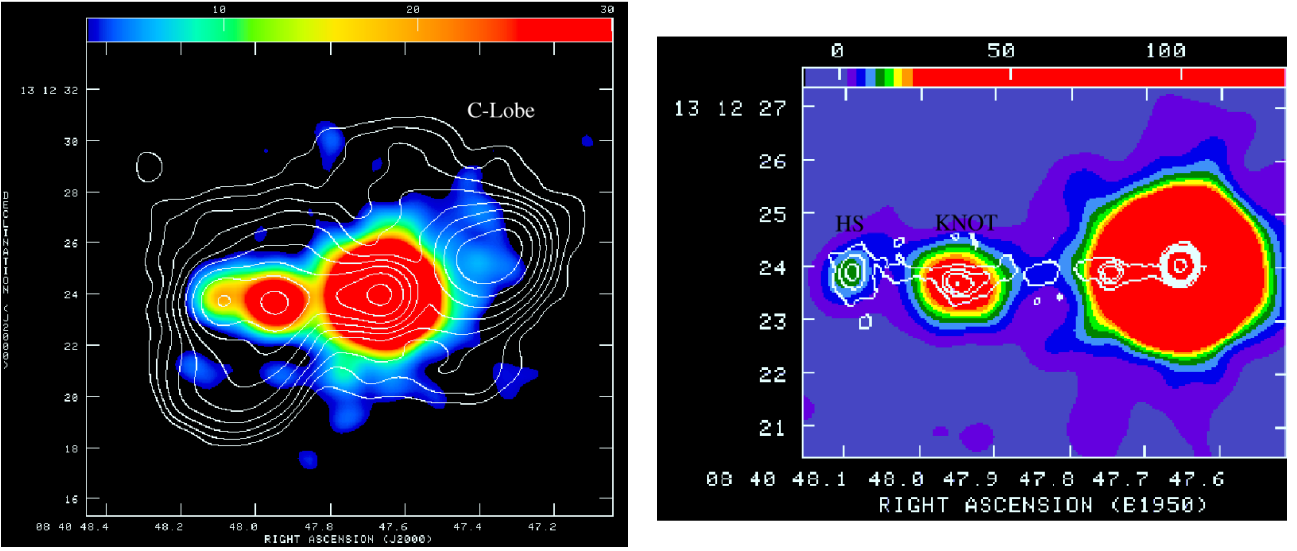}}
\caption[]{The {\it Chandra} 0.2--10 keV image
(color) is superimposed on the VLA 1.4 and
15 GHz isophotes ({\bf left} and {\bf right} panel respectively).
A $\simeq 0.9$ arcsec shift has been applied to match the X--ray and radio
peaks.
{\bf Left Panel}:
The {\it Chandra} image is obtained with a smoothing ($\sigma$=1.8 pixels)
of a subsampled (0.25 arcsec square pixels, i.e. binned at 0.5 original
pixels) image.
The 1.4 GHz radio contours are: 
$1 \cdot 10^{-3} \times$ (-1, 1, 2, 4, 8, ...) Jy/beam;
the FWHM of the beam is 1.4$\times$1.4 arcsec
and the flux peak in the image is 0.315 Jy/beam; this resolution is
comparable to that of the X--ray image.
{\bf Right Panel}:
Close up of the X--ray and radio jet.
The {\it Chandra} image is obtained with a smoothing ($\sigma$=2 pixels)
of a subsampled (0.1 arcsec square pixels, i.e. binned at 0.2 original
pixels) image.
The 15 GHz radio contours are: $8 \cdot 10^{-4} \times$
(-1, 1, 4, 8, 16..) Jy/beam; the FWHM of the 
beam is 0.25$\times$0.25 arcsec
and the flux peak in the image is 0.757 Jy/beam.}
\end{figure*}

\begin{table*}
\caption{X--ray and radio spectral analysis}
\begin{tabular}{cccccccccc}
\hline
Component   & $N_{\rm H}$  & $\Gamma$   & Norm & $\chi^2$/dof & kT 
& $\chi^2$/dof & $\alpha_{1.4}^{4.8}$ & $\alpha_{4.8}^{8.6}$ &
$\alpha_{8.6}^{14.9}$ \\
 & ($10^{20}$cm$^{-2}$) & & (${{10^{-6} {\rm ph}}\over 
{{\rm cm}^2\, {\rm s} \, {\rm keV} }}$) & & (keV)& & & & \\
\hline
Nucleus  & 16.5$\pm$3.5 & $1.22_{-0.05}^{+0.06}$   
& $163_{-12}^{+9}$ & 201.2/182 & ... & ... & ... & ... & ...\\
Knot   & 5.4 (f)& 1.23$^{+0.33}_{-0.29}$ & 4.6$\pm$0.1 & 11.4/10 & 
 80($>$10.4) & 
11.7/10 & 0.78$\pm$0.06 & 0.88$\pm$0.12 & 1.83(1.09)$\pm$0.13\\
Hot Spot  & 5.4 (f)& 1.7$^{+1.3}_{-0.7}$  & 
1.3$\pm$0.5 & ...  & ...   &  ... &
0.80$\pm$0.06  & 0.88$\pm$0.12 & 1.43(1.17)$\pm$0.13 \\
C--lobe & 5.4 (f) 
& 1.46$^{+0.36}_{-0.34}$ & 4.5$\pm$0.1 & 9.6/9 & 15($>$5.7) 
& 9.4/9 & 0.92$\pm$0.06 & 0.89$\pm$0.12 & ...\\
\hline
\end{tabular}
\end{table*}
          
Standard data reduction was done using the National Radio Astronomy 
Observatory (NRAO) AIPS package. We used the 1.4, the 4.8 and 8.4
GHz B array data to obtain ``low resolution'' images with a circular 
restoring beam of 1.4 arcsec. The self-calibration and imaging procedure
at each frequency was performed using only the uv-points within a common
minimum and maximum baseline.
These images allow to derive morphological and spectral
information of the main jet and of the extended lobe emission.
The 8.4 GHz A array and the 14.9 GHz data were used
to obtain 
``high resolution'' images with a circular restoring beam of 0.25
arcsec. The self-calibration and imaging procedure at each frequency
were performed using only the uv-points within a common minimum and maximum
baseline. 
The ``high resolution'' image (Fig.1) resolves the radio jet 
in three main components: an innermost knot at $\sim$2 arcsec from the
nucleus, a second knot at $\sim$4 arcsec from the nucleus and a
hot spot at the end of the jet.
Since {\it Chandra} cannot separate the innermost knot 
from the luminous nuclear source, in the following we focalize
on the knot at 4 arcsec distance and on the hot spot.  

The radio spectra of the knot and of the hot spot are relatively
steep with a similar spectral index ($\alpha \sim 0.8-0.9$,
see Tab.2).
Although higher frequencies ($\geq 100$ GHz) radio observations  
would be necessary to better describe the spectrum of both 
knot and hot spot, 
we find some evidence of a high frequency spectral steepening.
From an inspection of the radio fluxes 
(UMRAO Database Interface, 
http://www.astro.lsa.umich.edu/obs/radiotel/umrao.html) at the epochs of 
the archive data (Tab.1) 
we find that 3C 207 shows moderate flux variability at a level of
$\leq 40\%$ at 15 GHz, probably induced by the nuclear component, whereas
the variability at lower frequencies (e.g., 4.8 GHz), 
where the contribution of the nuclear component 
is expected to be slightly lower, is reduced to $\leq 20$\%.
This evidence, combined with the lack of synchrotron self
absorption in the spectra of both knot and hot spot, 
suggests that for these components
the spectral variability is not important
and that the derived
spectral indices (Tab.2) are robust.

\begin{figure}
\includegraphics{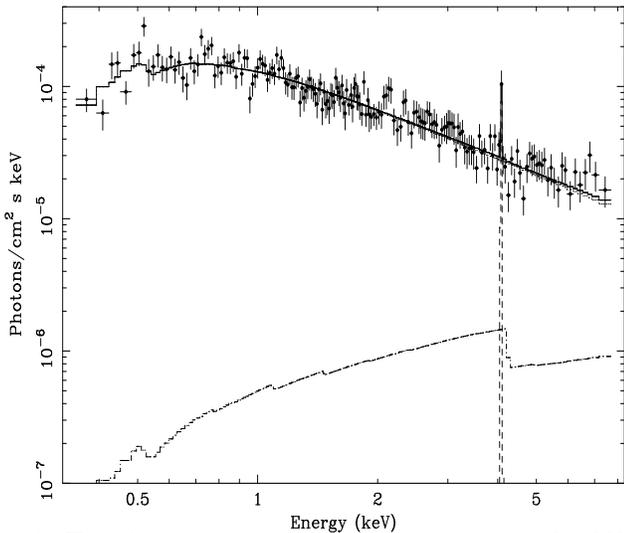}
\vspace{7 cm}
\caption[]{The {\it Chandra} spectrum of the nucleus of 3C 207.
The fitted model is a power law ($\Gamma=1.22$) component
absorbed by a column density  
$N_{\rm H}=1.65\times 10^{21}$cm$^{-2}$, plus a narrow gaussian
iron line component centered 
at 6.87 keV (rest frame) with a rest frame 
equivalent width of 150 eV (dashed).
A reflection component from a Seyfert--like spectrum
(dot--dashed) is also added to the model (see text).} 
\end{figure}

A visual inspection of the 1.4--8.4 GHz spectral index image 
of the western lobe does not show clear spatial trends 
(steepening or flattening) about the average value 
$\alpha \sim 0.9$, whereas at higher frequencies 
it is difficult to obtain spectral information due to the
lack of the adequate short baselines.
This may indicate a rather constant electron spectrum ($\delta \sim
2.8$; $N(\gamma) \propto \gamma^{-\delta}$) in the radio volume.
However, as 3C 207 is classified as a quasar,
its radio axis is seen at a relatively small angle
with the line of sight and projection effects may be 
important. A relatively constant spectral index 
in the lobes might derive from the  
mixing of regions with flatter and steeper spectrum intercepted
by the line of sight.
As in the case of the global (i.e. spatially unresolved)
spectrum of radio galaxies, the mixing of different spectra
in 3C 207 can be fitted with a synchrotron 
continuous injection model (e.g., Kardashev 1962). 
We find that a continuous injection model 
with a flat injected spectrum, $\delta \sim 2$, fits the
spectrum of 3C 207 if the break energy (i.e. the highest 
energy of the oldest electrons in the radio volume) 
is $\gamma_{\rm b} < 2 \times 10^3$ 
(assuming typical magnetic field strengths).

\subsection{Chandra Data}

The quasar 
3C 207 was observed for 37.5 ksec with the {\it Chandra} X--ray observatory 
on 2000 November 4. The raw level 1 data were re--processed using the latest 
version (CIAO2.1) of the CXCDS software.
The target was placed about 40'' from the nominal aimpoint 
of the chip 7 (S3) in ACIS--S and thus the on--axis point spread 
function applies. 
The lightcurve of the full chip shows evidence for particle background flares 
in the detector during the last part of the observation. The high background
times were filtered out leaving about 30 ksec of useful data
which were used in the spectral analysis described below.

Inspection of the X--ray image (Fig.1)
shows several immediately apparent features: 
a bright pointlike source coincident with the radio nucleus, 
enhanced diffuse emission in the direction of the western 
counter--lobe (C--lobe) 
and extended emission in the direction of the radio jet concentrated in two
relatively bright X--ray knots.
These X--ray knots are spatially coincident with  
one of the radio knots and with the eastern hot spot. 
X--ray spectra have been extracted 
using appropriate response and effective area functions.
The spectral data have been grouped into bins with a minimum of 20 counts
for the nucleus and 15 counts for the X--ray knot associated with the 
innermost part of the jet and for the extended emission in the opposite 
direction. We also examined the spectrum of the X--ray hot spot    
but with about 40 counts we can only obtain rough spectral
information.
The nuclear spectrum is well fitted by a single absorbed power law
(Fig.2, Tab.2).
The excellent counting statistic allowed to obtain stringent constraints
on the spectral parameters. The power law slope is extremely flat
$\Gamma$ = 1.22$\pm$0.06 while the best fit absorption column density 
is significantly higher than the Galactic value in the direction 
of 3C 207 ($N_{\rm H}$ = 5.4 $\times$ 10$^{20}$ cm$^{-2}$, 
Elvis et al. 1989). Fixing the column density at the 
Galactic value and adding an intrinsic absorber at the source redshift 
the best fit $N_{\rm H}$ value is 
(1.65$\pm$0.35) $\times$ 10$^{21}$ cm$^{-2}$.
There is also evidence of excess emission around 4 keV. 
The addition of a narrow gaussian iron line improves the fit
at the 98\% confidence level (according to the F--test).
It is interesting to note that the best fit line energy 
of 6.87$\pm$0.05 keV in the rest frame 
(the error is at the 90\% confidence
level for one parameter) 
strongly suggest the presence of a highly ionized gas 
in the nucleus of 3C 207. 
The observed equivalent width of 91$\pm$50 eV
corresponds to 153$\pm$84 eV rest frame.
Relatively strong iron lines are common among Seyfert galaxies and
and low luminosity radio--quiet quasars while they are rarely
observed in radio loud objects (Cappi et al. 1997).
A likely explanation is that the beamed non--thermal component
overshines the accretion disk emission.
In order to model the Seyfert--like and beamed non--thermal
contributions 
a reflection component from a Seyfert--like steep power law
(fixed at $\Gamma$=1.9) plus a narrow Gaussian iron line have been added
to the flat power law non--thermal spectrum.
Although poorly constrained we obtained a good
fit to the data with a reflection normalization R $\simeq$ 0.85
(where R= $\Omega$/2$\pi$ and $\Omega$ is the solid angle irradiated
by the X--ray source).
It is interesting to note that similar results have been recently reported
by Reeves et al. (2001)
from Newton--XMM observations of the high redshift ($z=3.104$)
radio--loud quasar PKS 0537$-$286.
In their case, however, the iron line strength implies a much weaker
reflection component (R $\simeq$ 0.2) possibly indicating 
a larger non--thermal beamed contribution as also indicated  
by the very flat radio synchrotron spectrum of this object. 

\begin{figure}
\includegraphics{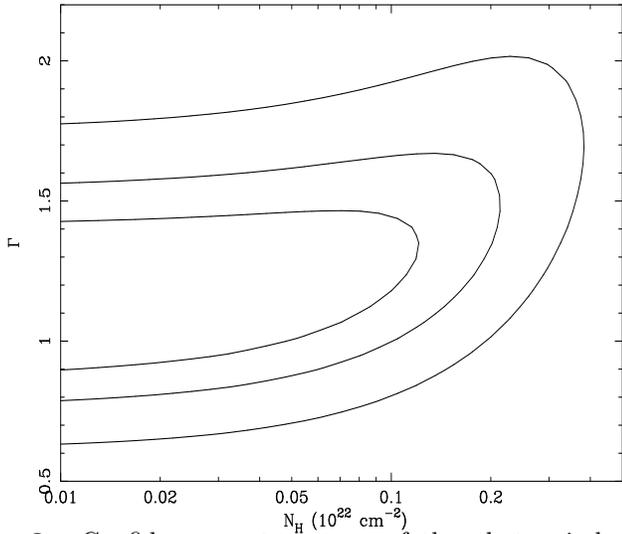}
\vspace{7 cm}
\caption[]{
Confidence contour map pf the photon index, $\Gamma$, as
a function of 
the column density, $N_{\rm H}$, in the case of the knot.
The reported contours are at 68, 90 and 99 \% confidence
level. The Galactic column density is $5.4 \times 10^{20}$cm$^{-2}$.}
\end{figure}

\begin{figure}
\includegraphics{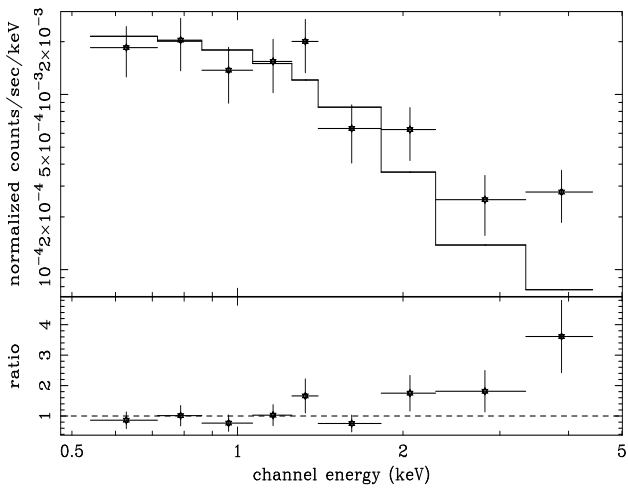}
\vspace{7 cm}
\caption[]{The {\it Chandra} 0.5--5 kev spectrum of the 
extended emission projected on the western radio lobe
(C--lobe)
is compared with a thermal model with 0.3 solar metallicity
and ${\rm kT}$=4 keV.} 
\end{figure}

The spectral parameters of the non--nuclear extended components are reported 
in Table 2. 
In all the cases a relatively flat power law, with the absorption
fixed at the Galactic value, does provide
an acceptable description of the data.
We have evaluated the influence of background subtraction
on the derived spectral parameters using different extraction regions
over the detector. The {\it Chandra} background is always extremely
low and has a negligible effect on the spectral parameter estimation.
Their uncertainty is dominated by the 
counting statistics; in Fig.3 we report 
the confidence contour map of the photon index
and column density in the case of the knot.
An equally good description of the data, 
from the statistical point of view,
is obtained using a thermal model with abundances fixed at 0.3 solar. 
It is important to point out that although the quality of the data
is not such to discriminate between the two options 
(power--law and thermal)
the derived best fit temperatures are 15 keV (${\rm kT} > 5.7$ keV 
at 90\% conf. level; Fig.4)
and 80 keV (${\rm kT} > 10.4$ keV at 90\% conf. level) for 
the C--lobe and the bright knot respectively.
The relatively high temperature requested to fit the 
C--lobe spectrum excludes the possibility that a 
cooling flow from a surrounding low brightness X--ray cluster
can significantly contribute to this emission.
Alternatively, one may wonder that the C--lobe emission
is the core region of a surrounding X--ray cluster.
From the luminosity--temperature correlation
(e.g., Arnaud \& Evrard 1999) one has that 
thermal emission from a hot cluster (with ${\rm kT} > 6$ keV)  
would provide a luminosity $> 10^{45}$erg s$^{-1}$.
However, we find that the 0.1--10 keV luminosity in excess to the 
core and jet emission (calculated in a circular region of 200 kpc
radius) is only $\sim 8-10 \times 10^{43}$erg s$^{-1}$
depending on the assumed X--ray spectrum and background.
In addition most of this luminosity 
($\sim 6-7 \times 10^{43}$erg s$^{-1}$) 
is produced in the C--lobe region where
the morphology of the X--ray counts is similar 
to the radio brightness distribution. 
We conclude that although the presence of a low brightness
surrounding cluster cannot be excluded, 
the great majority of the C--lobe X--ray counts
are emitted by a non--thermal 
process.

\section{Interpretation of the data}

As discussed in the previous Section the {\it Chandra}
spectra and morphology of both the lobe--like and
jet--like features are not consistent 
with a thermal scenario. 
An important finding is that 
the radio and the X--ray spectra are different, 
with the X--ray spectra 
sistematically flatter than the radio ones.
This is not in contrast with a non--thermal scenario.
Indeed, X--rays produced by IC or SSC processes 
may select portions of the relativistic electron
spectrum quite different from those responsible
of the synchrotron radio emission.

In order to explain the observed X--ray properties of 3C 207
we have worked out the spectrum of the relativistic
electrons under the very simple assumptions of the standard
scenario in which the particles are accelerated/reaccelerated
by strong shocks taking place in the compact
regions of the radio sources, i.e.
hot spots and knots in the radio jets 
(e.g., Meisenheimer et al., 1997 and ref. therein).

In the framework of shock acceleration theory 
only particles with Larmor radii larger than
the thickness of the shock are actually able to feel the 
discontinuity at the shock. The shock thickness is of the order of 
the Larmor radius of thermal protons, so that a cut--off at 
lower energies is formed in the accelerated spectrum 
(e.g. Eilek \& Hughes 1990, and ref. therein).
The presence of
such a cut--off should be taken into account when low energy electrons 
contribute to the emitted X--ray spectra via the IC process.
The presence of Coulomb losses would further contribute to
the flattening of the electron spectra at lower energies
(e.g., Sarazin 1999 and ref. therein), but here they are not
considered for simplicity as the basic picture would not
change.

The readers not interested in the details can skip sub--section 3.1 
(referring to Fig.5) and go
directly to sub--sections 3.2 and 3.3 where the expected lobe and jet 
emissions are discussed.

\subsection{The spectrum of the relativistic electrons}

We assume a simplified two steps scenario :

a) relativistic electrons, continuously
injected in the knots and hot spots with a power law spectrum,  
are accelerated/re--accelerated in the
shock regions subject to radiative
losses, thus the final spectrum is given by the competition 
between acceleration and loss mechanisms; 

b) the accelerated electrons diffuse from the shock regions
and continuously fill a post--shock region 
where older electrons are mixed with those more recently injected.
In this region electrons are subject to radiative losses only.

Following Kirk et al.(1998), 
we assume that electrons are injected and reaccelerated
by Fermi I -- like processes in the shock region from which
they typically escape in a time $T_{\rm es}$.

The particle energy
distribution,
${\cal N}(\gamma)$, can be obtained 
by solving the equation of continuity (e.g. Kardashev 1962)
with a time--independent approach :

\begin{equation}
{d \over{d \gamma}} \left[ {\cal N}(\gamma) \right]
{{d\gamma}\over{dt}}
+ {\cal N}(\gamma) \left[
{{d^2 \gamma}\over{d\gamma dt}}
+ {1\over{T_{\rm es}}} \right]
-{{ \rm Q}}(\gamma) = 0
\label{kinetic_0}
\end{equation}

\noindent
where $d \gamma/ d t =-\beta \gamma^2 + \chi_{\rm a} \gamma$
($\chi_{\rm a}$ and $\beta$ the acceleration 
and radiative losses coefficient, respectively) and
$Q$ is the particle injection rate.
The resulting total electron spectrum for a continuous
injection represented by a truncated power law
${{\rm Q}}(\gamma) = {{\rm Q}} \gamma^{-s}$
($\gamma < \gamma_*$) is :

\begin{equation}
{\cal N}(\gamma)=
{{{{\rm Q}}}\over{\chi_{\rm a}}}
\left( 1- {{\gamma}\over{\gamma_{\rm c}}} \right)^{-\alpha_-}
\gamma^{-\alpha_+}
\int_{\gamma_{\rm low}}^{\tilde{ \gamma}}
{{ x^{\alpha_+-(1+s)} }\over{
\left[ 1- {{x}\over{\gamma_{\rm c}}} \right]^{1-\alpha_- } }}
d x
\label{n_as}
\end{equation}

\noindent
where $\gamma_{\rm low}$ is the minimum energy of the electrons
that can be accelerated by the shock, 
$\tilde{\gamma}=\min \left( \gamma_* \, \gamma \right)$, 
$\alpha_+=1+[\chi_{\rm a} T_{\rm es}]^{-1}$,  
$\alpha_-=1-[\chi_{\rm a} T_{\rm es} ]^{-1}$,
and the high energy cut--off 
$\gamma_{\rm c} = \chi_{\rm a} / \beta$
is given by the ratio between gain and loss terms.
We notice that the slope of the electron spectrum 
for $\gamma > \gamma_*$ does not depend on the
spectrum of the electrons injected in the shock.
In the following we restrict ourself to the case
$s=\alpha_+$ and  
$\chi_{\rm a} \sim T_{\rm es}^{-1}$ so that $\alpha_-=0$,  
$\alpha_+=2$ and the spectral shape is that provided 
by the classical strong diffusive shock acceleration
(e.g. Heavens \& Meisenheimer 1987, and ref. therein).

The post--shock region 
is continuously supplied by the electron spectrum 
given by Eq.\ref{n_as} whose time evolution 
due to radiative losses (an approximatively 
constant magnetic field strength $B$ is assumed) 
is obtained by solving the time--dependent continuity equation :

\begin{equation}
{{\partial {{\cal N }}(\gamma, t)}\over
{\partial t}} =
-{{\partial}\over{\partial \gamma}} 
\left[ {{d \gamma}\over{d t}} {{\cal N }}(\gamma,t) \right]
\end{equation}

\noindent
so that 

\begin{equation}
{{\cal N }}(\gamma,t)=
{{{{\rm Q}} \,\, {{\rm C}} }\over{\chi_{\rm a}}}
\gamma^{-2} \int_{\gamma_{\rm low}}^{{\tilde{\gamma}(t)}}
x^{-1}
\left[ 1- {{x}\over{\gamma_{\rm c}}} \right]^{-1}
d x
\label{n(t)}
\end{equation}

\noindent
where ${\rm C} \sim 1/T_{\rm es}$ provides for the
conservation of the flux of the electron number across the shock region, 
and 

\begin{equation}
\tilde{\gamma}(t)=
{\rm min} \left\{ \gamma_* \,\, , \,\, {{\gamma}\over{1 - \gamma 
\beta t}}
\right\}
\label{extr}
\end{equation}

\begin{figure}
\resizebox{\hsize}{!}{\includegraphics{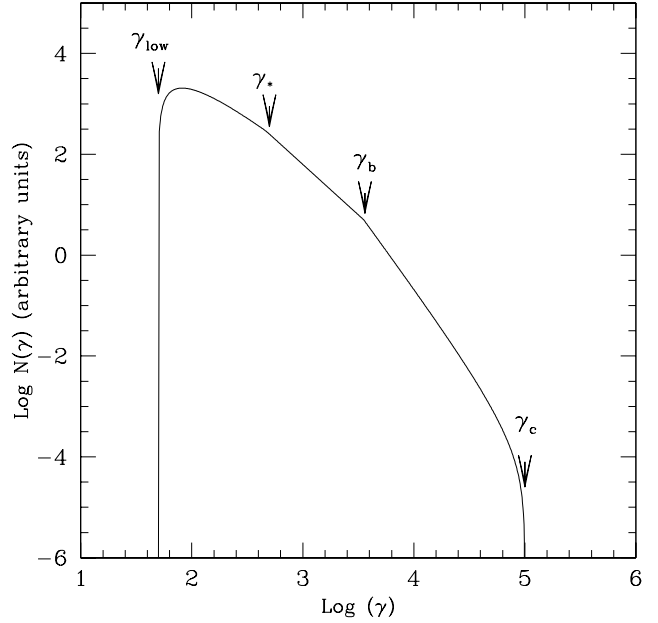}}
\caption[]{A representative spectrum of the accelerated 
relativistic electrons is reported with the relevant parameters
indicated in the panel :   
the largest energy of the electrons injected in the post--shock
region ($\gamma_{\rm c}$), the largest energy of the oldest electrons
in the considered volume ($\gamma_{\rm b}$), 
the largest energy of the electrons injected
in the shock region ($\gamma_{*}$), and 
the lower energy of the electrons accelerated by the shock
($\gamma_{\rm low}$).} 
\end{figure}

\noindent
Finally, in a post--shock region with size  
determined by the diffusion length of the particles in the
largest considered time T, 
the volume integrated spectrum of the electron population, 
$N(\gamma)$, is given by the sum of all the injected
electron spectra, i.e. it is obtained 
by integrating Eq.\ref{n(t)} over the time interval 0--T.  
The solution
is given by the following three cases covering different
portions of the energy spectrum : 

\noindent
$\bullet$
$\gamma > \gamma_{*}$

\begin{equation}
N(\gamma)=
{{ {\rm Q} \, C }\over{ \chi_{\rm a} \beta }}
\gamma^{-2} {\rm ln} \left(
{{ 1- \gamma_{\rm low}/\gamma_{\rm c} }\over{
1- \gamma_{*}/\gamma_{\rm c} }} \times
{{ \gamma_{*} }\over{ \gamma_{\rm low} }} \right)
T_{\gamma}
\label{n_fin1}
\end{equation}

\noindent
$\bullet$ $\gamma< \gamma_{*}$ and $T_{\gamma} >  
\gamma^{-1}
(1-\gamma/\gamma_{*})$ 

\begin{eqnarray}
N(\gamma)=
{{ {\rm Q} \, C  }\over{ \chi_{\rm a} \beta }}
\gamma^{-2} \times \{
{\rm ln} \left( {{ 1- \gamma_{\rm low}/\gamma_{\rm c} }\over{
1- \gamma_{*}/\gamma_{\rm c} }} \times 
{{ \gamma_{*} }\over{ \gamma_{\rm low} }} \right) T_{\gamma} + \nonumber\\
{ 1\over {\gamma  }}
(1 - {{\gamma}\over{\gamma_{\rm c}}} )
\left[ {\rm ln}\left( {{\gamma}\over{\gamma_{*} }}
-{{\gamma}\over{\gamma_{\rm c}}} \right)
- {\rm ln}\left( 1 - {{\gamma}\over{\gamma_{\rm c}}} \right)
\right] + \nonumber\\
{ 1\over {\gamma  }} \left(1 - {{\gamma}\over{ 
\gamma_{*}}}
\right) \}
\label{n_fin2}
\end{eqnarray}

\noindent
$\bullet$ $\gamma< \gamma_{*}$ and $T_{\gamma} < 
\gamma^{-1}
(1-\gamma/\gamma_{*})$

\begin{eqnarray}
N(\gamma)=
{{ {\rm Q} \, C }\over{ \chi_{\rm a} \beta }}
\gamma^{-2} \times
\{ 
T_{\gamma} \left[
1 + {\rm ln} \left( {{ \gamma_{\rm c}/\gamma_{\rm low} -1  }\over{
 {{ \gamma_{\rm c} }\over{\gamma}} -1 
-\gamma_{\rm c} T_{\gamma}   }} 
\right) \right] +
\nonumber\\
{ 1\over {\gamma  }} \left(1 - 
{{\gamma}\over{\gamma_{\rm c}}} \right)
\left[
{\rm ln} \left( 1- {{ \gamma}\over{\gamma_{\rm c}}} -
\gamma  T_{\gamma} \right)
-{\rm ln} \left( 1- {{ \gamma}\over{\gamma_{\rm c}}} \right)
\right] \}
\label{n_fin3}
\end{eqnarray}

\noindent
where 

\begin{eqnarray}
T_{\gamma} =
\cases{
(\gamma_{\rm c}/\gamma_{\rm b} -1)/\gamma_{\rm c}
\, \,\,\,\,\,\, \,\,\,\,\,
\,\,{\rm for} \,\,\,\,\, 
\gamma < \gamma_{\rm b} \cr
(1- \gamma/\gamma_{\rm c})/\gamma 
\,\,\,\,\,\,\,\,\,\, {\rm for} \,\,\,\,\,
\gamma_{\rm b}
< \gamma < \gamma_{\rm c} \cr
0 \,\,\,\,\,\,\,\,\,\,\,\,
\,\,\,\,\,\,\,\,\,\,\,\,\,\,\,\,\,\,\,\,\,\,\,\,\,\,\,\,
{\rm for} 
\,\,\,\,\, \gamma > \gamma_{\rm c} \cr
}
\label{t_gamma}
\end{eqnarray}

\noindent
and $\gamma_{\rm b} = \gamma_{\rm c} / (1+ \gamma_{\rm c} \beta
T)$ is the highest energy of the oldest electrons in the volume being
considered.

\noindent
In Fig.5 we show a representative 
electron energy distribution
obtained from Eqs.(\ref{n_fin1}--\ref{n_fin3}).
The shape of the obtained spectrum depends on the energy region:
around each one of the relevant values of the electron 
energy ($\gamma_{\rm low}$, $\gamma_{*}$,
$\gamma_{\rm b}$ and $\gamma_{\rm c}$)
corresponds a break in the spectral distribution. 
Under the condition $\gamma_{\rm low} << \gamma_{*} 
<< \gamma_{\rm b}$, 
at intermediate energies, 
for $\gamma_{*}  < \gamma < \gamma_{\rm b}$,  
the slope approaches to $\delta \simeq 2$ (as in the standard
diffusive acceleration from strong shocks), at higher energies,    
for $\gamma > \gamma_{\rm b}$, it approaches to $\delta \simeq 3$
up to a sharp cut--off around $\gamma \sim \gamma_{\rm c}$ 
(as in the standard diffusive acceleration from strong shocks
including radiative losses).
In addition at lower energies, for $\gamma < \gamma_{*}$ 
(i.e. the largest injected energy), 
the spectrum gradually flattens ($\delta < 2$) and a low energy cut--off 
is formed around $\gamma \sim \gamma_{\rm low}$.

With increasing the size of the considered post--shock region 
the corresponding $T$ increases and
consequently, the electron break energy, $\gamma_{\rm b}$, 
shifts at lower energies. 
Instead, the value of the high energy
cut--off, $\gamma_{\rm c}$, which is fixed by 
the electron spectrum of the youngest electrons in the considered 
region is constant.

If portions of the post--shock region 
of a fixed size and at different distances
from the shock are considered, 
both the age of the youngest and oldest
electrons in the volumes increases and
consequently, both $\gamma_{\rm c}$ and $\gamma_{\rm b}$
shift at lower energies.
As a consequence, the electron energy distribution steepens 
as well as the corresponding synchrotron emitted spectrum, so
that the synchrotron brightness rapidly decreases and the knot
disappears. The decrease of the brightness with distance from the
shock would be further enhanced by the effect of adiabatic losses
which shift the electron spectrum at even lower energies.

\noindent
The basic idea is that, in the radio jet, after a 
cycle a)--b) electrons can reach a new shock region (knot or
hot spot) where they are injected and re--accelerated (a--b).
In principle, the electron spectrum as injected at 
each shock should be calculated taking into account the effect of 
all the shock regions previously crossed by the relativistic
electrons. 
For a detailed numerical study of a relatively
similar scenario 
we remind the reader to Micono et al.(1999).
Here, for seek of simplicity, we assume that the electron spectrum as
injected at the shock is independent on the previous history
of the relativistic electrons so that Eqs.(\ref{n_fin1}--\ref{n_fin3})
provide the shape of the electron energy distribution in all the 
knots and hot spots of 3C 207.

\subsection{The C--lobe emission}

As discussed in Section 2.3 the {\it Chandra} observation
of 3C 207 has evidenced the presence of diffuse, 
most likely non--thermal X--ray emission in the western 
and more distant radio lobe (C--lobe). 

If the radio synchrotron spectrum 
($\alpha_{\rm r} \sim 0.9$, Sect.2.2) of the
lobe is extrapolated in the X--ray energy band, 
the {\it Chandra} observed flux
would be overestimated by about one order of magnitude.
This implies a steepening of the synchrotron spectrum 
at higher frequencies whereas the 0.5--2 keV spectral
shape ($\alpha_X= \Gamma-1 \sim 0.4$, Tab.2) is flatter
than the radio spectrum.
As a consequence, synchrotron emission from the 
C--lobe cannot significantly contribute to the 
observed X--ray flux.

The remaining non--thermal processes to be 
considered are: the IC scattering
of CMB photons and of the nuclear photons, whereas,
as it is well known, the SSC process cannot contribute
a significant X--ray flux in the case of extended radio 
sources.
The IC scattering of nuclear photons appears to be
the most promising process. 
That this might be the case is indicated by the inspection
of the Fig.6 where we have assembled the relevant
photon energy densities ($\omega_{\rm ph}$), averaged over the radio volume
sectors intercepted by the line of sight, as a function of the 
projected distance from the nucleus and for different inclination
angles ($\theta_{\rm i}$) of the radio axis.
For the nuclear source we estimate an optical to far IR isotropic
luminosity $\sim 4 \times 10^{46}$erg s$^{-1}$, based on the
60$\mu$m IRAS flux (Van Bemmel et al. 1998) and on 
the average spectral shape of the radio loud quasars given by
Sanders et al.(1989).
It is immediately clear that at an angular distance from the
nucleus between 3--5 arcsec, where extended X--ray emission is detected,
the radiation field is largely dominated by the quasar IR emission
for inclination angle $\theta_{\rm i} \geq 10^o$, 
typical of steep spectrum radio loud quasars.
Moreover, the IC scattering of the
CMB photons cannot explain the observed X--ray properties
for three main reasons:

\noindent
$\bullet$
contrary to the C--lobe no diffuse X--ray emission is observed 
in the eastern lobe despite the fact that the two radio lobes
have comparable sizes and radio luminosities;

\noindent
$\bullet$
the X--ray spectrum is much flatter ($\Delta \alpha \sim 0.5$, Tab.2)
than the synchrotron radio spectrum despite the fact that relativistic 
electrons of about the same energies are involved in the two
emission processes;

\noindent
$\bullet$
the estimated X--ray flux would be only $\sim 8\%$
of the observed flux unless the average magnetic field strength of 
the C--lobe is very much weaker than the equipartition value.

The basic requisite for the actual production of X--rays via
IC scattering of far--IR to optical nuclear photons
is that there are relativistic electrons of
sufficiently low energy in the lobes.
Then the observed properties of the extended emission from this
source can be readily explained:

\noindent
$\bullet$
the asymmetry in the lobes' emission is due to the dominance of the IC
back scattering in the far lobe, the C--lobe;

\noindent
$\bullet$
the slope of the X--ray spectrum reflects the flatter
slope of the low energy electrons compared with the radio
synchrotron electrons (Fig.5).

\begin{figure}
\resizebox{\hsize}{!}{\includegraphics{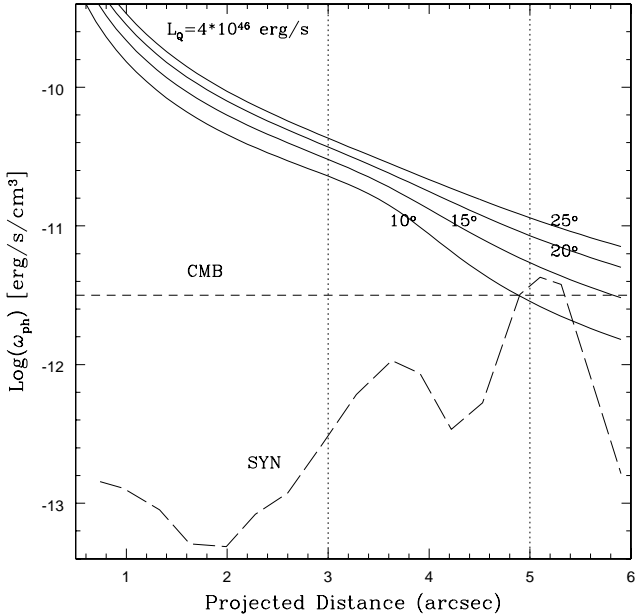}}
\caption[]{The energy density of the nuclear photons
(solid lines), averaged over the radio volume intercepted by the 
line of sight, is reported as a function of the projected 
distance from the nucleus for different angles between the
radio axis and the line of sight as given in the panel.
The energy densities of the synchrotron (long--dashed line)
and of the CMB (dashed line) are also reported.
Note that the synchrotron line between 4--6 arcsec
is not representative of energy density averaged over the
radio volume intercepted by the line of sight as it is 
calculated for the most compact radio features in the 
C--lobe.
The two vertical dotted lines mark the range of 
projected distance from the nucleus 
in which extended emission is detected.} 
\end{figure}

\begin{figure}
\resizebox{\hsize}{!}{\includegraphics{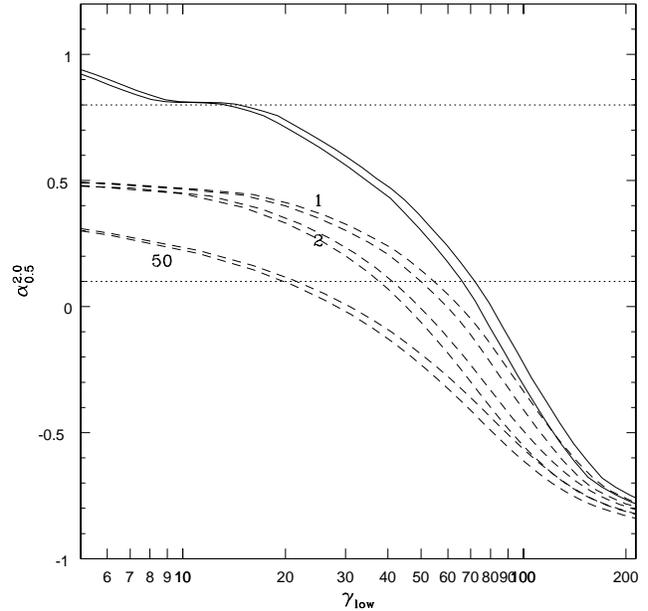}}
\caption[]{Expected 0.5--2 keV spectral indices as a 
function of the low energy cut--off ($\gamma_{\rm low}$).
The calculations are performed for different assumed
spectra of the emitting electrons:
a power law ($\delta=2.8$) extending down to $\gamma_{\rm low}$
(solid lines) and an accelerated electron spectrum 
(dashed lines) with $\gamma_*/\gamma_{\rm low}$= 1, 2, 50
as indicated in the panel.
For each model the lower and upper lines represent the values of the 
spectral index obtained for angles between the radio axis and the
line of sight $\theta_{\rm i}$= 10 and 30 deg, respectively.
The horizontal dotted lines give the observed 0.5--2 keV spectral index
interval at 90\% confidence level.} 
\end{figure}

\begin{figure}
\resizebox{\hsize}{!}{\includegraphics{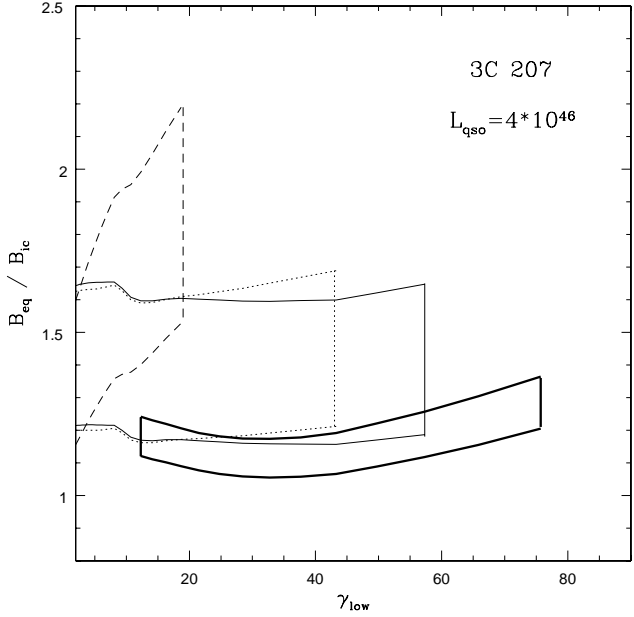}}
\caption[]{The allowance regions for the ratio between
IC--magnetic field and equipartition field strength 
and for $\gamma_{\rm low}$ are reported for different assumed
energy distributions of the electrons:
power law $\delta=2.8$ down to $\gamma_{\rm low}$ (thick solid
region), accelerated spectrum with $\gamma_*/\gamma_{\rm low}$=1
(solid region), =2
(dotted region) and =50
(dashed region).
In the case of reaccelerated electron spectra, the 
regions are calculated by assuming $\gamma_{\rm b}$ in the range
500--1000 and $\gamma_c >> 1000$.} 
\end{figure}

In order to obtain the expected X--ray emission
we use the anisotropic IC formulae given in Brunetti (2000),
integrated over the electron spectrum, nuclear photon energy
distribution and radio volume.

\noindent
We assume the electron energy 
distribution as given by Eqs.(\ref{n_fin1}--\ref{n_fin3}).
Although this spectral shape is calculated for the post--shock regions, 
it is still appropriate 
for the lobes if no efficient in situ reacceleration mechanisms are 
active in the lobe volume. 
The spectral parameters 
($\gamma_{\rm low}$, $\gamma_{\rm *}$, $\gamma_{\rm b}$ and 
$\gamma_{\rm c}$) 
are those of the post--shock region modified by  
adiabatic and  radiative
losses suffered in the lobes.

The match of the radio spectral index for a typical
equipartition field $B \sim 50 \mu$G requires $\gamma_{\rm b} \leq 10^3$
and $\gamma_* < \gamma_{\rm b}$.
With these constraints the 0.5--2 keV spectral index as a 
function of the combined values ($\gamma_{\rm low}$,
$\gamma_{*}$) is reported in Fig.7 ($\gamma_{\rm b} \geq 300$
is assumed).
It is seen that consistency with the {\it Chandra} data 
is achieved for $\gamma_{\rm low} < 60$.
For sake of completeness in Fig.7 we also report the result
of a simplified scenario in which the electron energy distribution
is obtained by simply extrapolating the spectrum of the 
radio synchrotron electrons ($\delta \sim 2.8$) down to an artificially
imposed sharp low energy cut--off $\gamma_{\rm low}$
(or similarly by releasing the assumption $\gamma_{\rm b} \geq 300$); 
in this case the fit to the {\it Chandra} data is obtained with 
$\gamma_{\rm low}=40 \pm 30$.

We stress that the main result, i.e. the extension of the electron 
energy distribution down to $\gamma$ of several tens, is 
robust as it does not depend on the details of the assumed 
electron spectrum.

The origin of the observed X--ray flux from  
the IC scattering of nuclear photons can
be used (Brunetti et al.1997)
to estimate the magnetic field strength in the radio
lobes and to test the minimum energy argument (equipartition).
In Fig.8 we report the allowed regions of the values of the
magnetic field strength (in units of the equipartition field)
and of $\gamma_{\rm low}$ as inferred by
the combined radio and X--ray measurements.
Each region is bounded by the allowed interval
of $\gamma_{\rm low}$ and $\gamma_*$ 
and by the two values of the
inclination angle, $\theta_{\rm i}=10^o$ and $30^o$, as 
obtained in Fig.7.
The corresponding equipartition fields are obtained numerically
by minimizing the total energy in the lobe with respect to
the magnetic field (with proton/electron energy ratio =1).
In Fig.8 we also plot the simplified case of a power law
electron spectrum ($\delta=2.8$) down to a low energy 
cut--off $\gamma_{\rm low}$; in this case the corresponding
equipartition fields are computed by applying the equations
of Brunetti et al.(1997).
As a general result, it is seen that the magnetic field
strengths are lower, but within a factor of $\sim 2$,
from the equipartition values.

\subsection{The knot emission}

We compare the radio and X--ray data of the knot
with the predictions of two models:

\noindent
$\bullet$
a homogeneous sphere at rest
emitting X--rays via SSC (e.g., Marscher 1983).

\noindent
$\bullet$
following the Tavecchio et al.(2000) and Celotti et al.
(2001) model for the jet of PKS 0637$-$752, we consider 
a homogeneous sphere moving with a bulk
Lorentz factor $\Gamma_{\rm bulk}$ at an angle, $\theta_{\rm bulk}$,
with respect to the line of sight emitting X--rays via 
IC scattering of CMB photons
(external inverse Compton, EIC).

\begin{figure}
\resizebox{\hsize}{!}{\includegraphics{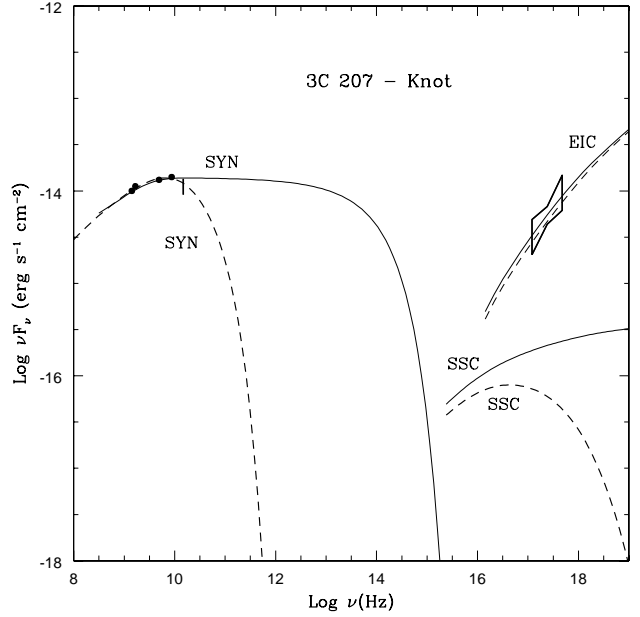}}
\caption[]{The radio and X--ray data of the main knot of 3C 207
are compared with the different model predictions.
Two representative synchrotron spectra are reported with 
the corresponding SSC spectra calculated by assuming the
equipartition
magnetic field ($\sim 280 \mu$G).
The corresponding EIC models are calculated by assuming 
$\Gamma_{\rm bulk}=6$, $\theta_{\rm bulk}=7^o$ and a value of the
magnetic field a factor of 1.5 lower than the equipartition field 
(knot rest frame).} 
\end{figure}

\noindent
Roughly speaking, in the case of boosting
(${\cal D}$, the Doppler factor), 
the electrons responsible for synchrotron radio emission have
energies :

\begin{equation}
\gamma_{\rm R} \sim 3 \cdot 10^4 
\sqrt{ {1 \over { B(\mu G) } {{\cal D}} }}
\label{er}
\end{equation}

\noindent
the minimum energy of 
those responsible for IC emission of CMB photons :

\begin{equation}
\gamma_{\rm EIC} \sim \left( {{ \nu_{\rm X}}\over{
\nu_{\rm CMB} }} \right)^{1\over 2}
{ {{\cal D}} }^{-1}
\sim 5 \cdot 10^2 { {{\cal D}} }^{-1}
\label{eic}
\end{equation}

\noindent
whereas, the energy of those responsible 
for SSC emission have energies: 

\begin{equation}
\gamma_{\rm SSC} \sim \left( {{ \nu_{\rm X}}\over{
\nu_{\rm R} }} \right)^{1\over 2}
\sim 10^4
\label{essc}
\end{equation}

For important boosting (${{\cal D}} \sim 10$)
one has $\gamma_{\rm SSC} \sim 10^4$, 
$\gamma_{\rm R} \sim  10^3$ ($B \sim 100 \mu$G) and
$\gamma_{\rm EIC} \sim 50$, whereas for no boosting or 
in the case of moderate de--boosting (i.e., 
$20^o < \theta_{\rm bulk} < 30^o$ and
$2< \Gamma_{\rm bulk} < 10$) one has
$\gamma_{\rm SSC} \sim \gamma_{\rm R} \sim 10^4$
and $\gamma_{\rm EIC} \sim 500$.

The most important result from the 
spectral analysis of the knot emission (Tab.2)
is the robust difference 
between the X--ray energy index ($\sim 0.2$) 
and the radio energy index ($\sim 0.8-0.9$) 
of the knot.
As the SSC X--ray flux is emitted by electrons with
energies similar to or greater than those of 
the radio electrons (Eqs.\ref{er} and \ref{essc}),  
this result would immediately rule out the possibility
that SSC process can power the X--ray flux of the knot.
In order to reproduce the data, 
the X--ray knot should be powered by IC scattering of electrons
at lower energies where the energy distribution is
flatter (Fig.5), as in the case of 
the EIC scattering model (Eq.\ref{eic}).
It should be remarked that the required spectral
flattening may also be obtained by a simple extrapolation
of the radio electron spectrum down to an imposed low
energy cut--off at $\gamma_{\rm low} \sim \gamma_{\rm EIC}$.
In order to perform a detailed model 
calculation, we assume the evolved
electron spectrum as given by 
Eqs.(\ref{n_fin1}--\ref{n_fin3}). 
The calculations of the synchrotron,
SSC and EIC spectra are performed with the equations 
given in the Appendices.
We calculate the value of the equipartition magnetic
field strengths by minimizing the electron 
energy density (as obtained by integrating 
Eqs.(\ref{n_fin1}--\ref{n_fin3})) with respect to the magnetic field 
for a given synchrotron spectrum; the energy ratio between
protons and electrons is taken =1.

The comparison between model predictions and data 
is reported in Fig.9.
The radio data provide marginal evidence
for a curvature in the synchrotron spectrum.
To test this high frequency radio observations and/or
deep HST optical measurements are necessary.
In order to fit the radio spectrum we obtain a break frequency 
(i.e. the emitted frequency corresponding to $\gamma_{\rm b}$)
$\nu_{\rm b} \sim 2-5$ GHz and a cut--off frequency
(i.e. the emitted frequency corresponding to $\gamma_{\rm c}$) 
$\nu_{\rm c} \geq 10$ GHz.
In terms of the Lorentz factors (with $\gamma_* < \gamma_{\rm b}$)
these constraints yield :

\begin{equation}
\gamma_{\rm b}^2 
B(\mu{\rm G}) {{\cal D}} 
\sim 0.7-1.8 \times 10^9
\label{gb}
\end{equation}

\noindent
and 

\begin{equation}
\gamma_{\rm c}^2 
B(\mu{\rm G}) {{\cal D}} 
\sim 5\times 10^5 \nu_{\rm c}({\rm MHz})
\geq 5\times 10^9
\label{gc}
\end{equation}

\noindent
Here, we limit ourself to reproduce the synchrotron
radio data with two extreme models, one with a cut--off at
high radio frequencies ($\sim 10^{10}$Hz)
and the other with a cut--off 
in the optical band.
In Fig.9 we report the SSC spectra corresponding
to the two mentioned cases.
It is seen that the SSC spectrum depends on the assumed 
synchrotron model while the EIC spectrum, 
which depends on the low energy
side of the electron distribution, is fairly independent on 
the assumptions about the high frequency cut--off. 

Both the X--ray flux 
and the spectral shape are well reproduced by an EIC model 
whereas a SSC model cannot account for the data.
As a matter of fact, if a value of the magnetic field strength
a factor of 6--7 times smaller than the equipartition
value is assumed, then the SSC model can reproduce the
observed X--ray flux but the derived X--ray spectral shape
is not consistent with that observed.
\begin{figure}
\resizebox{\hsize}{!}{\includegraphics{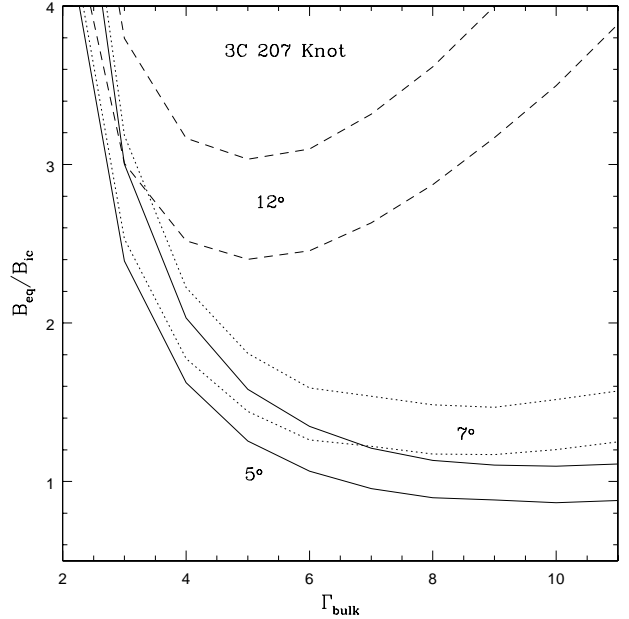}}
\caption[]{The ratio between the value of the equipartition
magnetic field and that
required by the EIC model to match the X--ray flux is reported
for the knot as a function of $\Gamma_{\rm bulk}$.
The calculations are performed 
for different values of $\theta_{\rm bulk}$ as reported in the panel;
the two lines reported for
each $\theta_{\rm bulk}$ give the allowed region for the  
value of the ratio obtained by taking $2 < \gamma_{\rm low} < 50$
and $100 < \gamma_* < 300$.
For each set of the parameters, 
the equipartition magnetic field strength is calculated
in the knot rest frame.} 
\end{figure}
In order to fit the the X--ray spectral shape with the EIC
model, the electrons responsible for the X--ray flux should have
a relatively flat energy distribution (Fig.5) so that 
$\gamma_{\rm low}$\ltsim$\gamma_{\rm EIC} < \gamma_*$ is required;
i.e., for substantial boosting it should be 
$\gamma_{\rm low}$\ltsim$50 < \gamma_*$.
  
In Fig.10 we report the ratio between the value of the 
equipartition magnetic field strength 
and that required by the EIC model to match the X--ray flux
as a function of $\Gamma_{\rm bulk}$ for different values
of $\theta_{\rm bulk}$.
According to the constraints on $\gamma_{\rm low}$ and 
$\gamma_*$ obtained from the X--ray spectrum, 
the calculations in Fig.10 are performed for 
$2<\gamma_{\rm low}<50$ and $100< \gamma_* < 300$.
By requiring a magnetic field strength close 
to (within a factor of 2) the equipartition value, 
we find that $\theta_{\rm bulk} \leq 10^o$ and
$\Gamma_{\rm bulk} \geq 4$ are necessary to match the data.
In this case, from Eqs.(\ref{gb}--\ref{gc}) and from the calculated
$B_{\rm ic}$ and ${\cal D}$ one typically has 
$\gamma_{\rm b} \sim 1400-2300$ and 
$\gamma_{\rm c} \geq 3800$.
For $\gamma_* > 300$ (but always $\gamma_* < \gamma_{\rm b} \sim 10^3$)
the EIC is less efficient and 
slightly larger departures from the equipartition
condition are required to match the X--ray flux.
Similarly, for $\gamma_{\rm low}$\ltsim$50$ the number of electrons
around $\gamma_{\rm EIC}$ is depleted and 
larger departure from the equipartition condition are required 
to match the X--ray flux.
In addition, we find that for $\gamma_{\rm low} > 100$
the derived X--ray spectrum is too hard and the EIC model 
cannot reproduce the data.

A less efficient boosting (e.g., 
higher values of $\theta_{\rm bulk}$ 
or smaller values of $\Gamma_{\rm bulk}$)
would require substantial
departures from the equipartition condition to match the X--ray
flux.
In this case slightly higher energetic electrons are involved
in the scattering and values of the low energy cut--off
$\gamma_{\rm low} > 100$ could still be consistent with
the X--ray spectral shape.

\subsection{The hot spot emission}

\begin{figure}
\resizebox{\hsize}{!}{\includegraphics{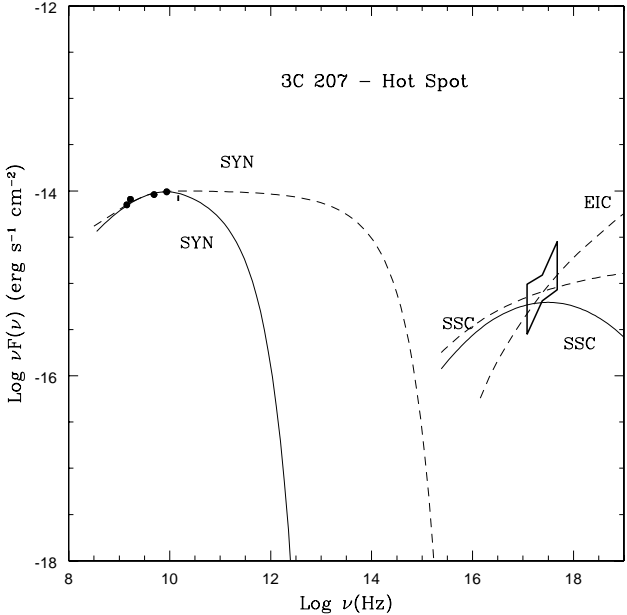}}
\caption[]{The radio and X--ray data of the eastern hot
spot of 3C 207 are compared with different model predictions.
Two representative synchrotron spectra are reported with 
the corresponding SSC spectra calculated by assuming a magnetic 
field value a factor of 2 lower than the equipartition 
($\sim 270 \mu$G).
The EIC model is calculated by assuming $\Gamma_{\rm bulk}=3$, 
$\theta_{\rm bulk}=10^o$ and a magnetic field value
a factor of 2 lower than the equipartition field 
(hot spot rest frame). } 
\end{figure}

We follow the same analysis procedure as outlined in the
preceeding Section 3.3 for the knot.
The radio and X--ray data are represented in Fig.11.
As for the knot there is a marginal evidence that the synchrotron
spectrum steepens at high radio frequencies
so that we have fitted the radio data with two different
synchrotron spectra, with the relevant parameters
consistent with those in the case of the knot.
Future high sensitivity high frequency observations
of the hot spot are needed to better constrain the
synchrotron spectrum.
The X--ray emission model of the hot spot is not well constrained
by the data.
As shown in Fig.11 the X--ray data are reproduced by both the
SSC model and EIC model with moderate Lorentz boosting 
($\Gamma_{\rm bulk} \sim 2-4$, and $\theta_{\rm bulk} \sim 5-10^o$).
The spectrum obtained with the EIC models provides a better fit
to the observed X--ray spectral shape, however the requirements 
of boosting is not strong. 
So far no kinematical evidence for relativistic motions 
of radio hot spots has been found. 
By assuming a value of the magnetic field in the hot spot 
within only a factor of $\sim 2$ 
smaller than the equipartition value
a homogeneous SSC model satisfactorily
reproduces both the observed X--ray flux and
the spectrum; the departure from equipartition is 
further reduced ($B_{\rm eq} \sim 1.7 \times B_{\rm ic}$)
if the contribution of the IC scattering of CMB photons is 
added to that of the SSC process.

\section{Discussion and conclusion}

The {\it Chandra} observation of the double lobed, steep spectrum 
radio loud quasar 3C 207 has revealed a strong nuclear source,
a one sided jet--like feature coincident with the eastern radio jet
and finally extended X--ray emission mostly associated with
the western radio counter lobe (C--lobe).
The jet--like emission is characterized by a luminous spatially
resolved knot and by X--ray emission coincident with the eastern radio
hot spot.

$\bullet$
The nuclear source (0.1--10 keV luminosity $\sim 3.2 \times 10^{45}$erg
s$^{-1}$) is detected with $\sim 5000$ net counts, the excellent count
statistics allows to perform a stringent spectral analysis.
The spectrum is well fitted by a flat power 
law component ($\Gamma=1.22^{+0.06}_{-0.05}$) absorbed by a column density 
$N_{\rm H}=1.65 \pm 0.35 \times 10^{21}$cm$^{-2}$, significantly
in excess to the Galactic value ($N_{\rm H}=0.54\times 10^{21}$cm$^{-2}$).
This result, confirms and improves the evidence for absorption in excess 
in this quasar ($N_{\rm H}=2.9^{+3.0}_{-2.15} \times 10^{21}$cm$^{-2}$) 
as found by Fiore et al.(1998) with ROSAT.
The spectral analysis has also revealed the presence of a ionized narrow 
iron line at $6.87\pm 0.05$ keV (source
frame) with intrinsic equivalent width of $153\pm 84$ eV.

$\bullet$
We have shown that the extended X--ray emission detected in the direction
of the western radio lobe (of 0.1--10 keV luminosity
$\sim 6-7\times 10^{43}$erg s$^{-1}$) 
is of non--thermal origin and can be best
interpreted as IC scattering of the IR photons from the quasar with 
relativistic electrons of much lower energies (typical Lorentz factors
$\gamma \sim 100$) than those responsible of the synchrotron radio
emission.

Accounting for the 0.5--2 keV spectrum leads to a robust upper limit,
$\gamma_{\rm low} < 70$, to the low energy cut--off of the
electron energy distribution.
This, together with a similar result obtained from the analysis
of the one sided X--ray lobe emission of the radio galaxy
3C 295 (Brunetti et al., 2001), 
indicates the presence of
substantial amounts of low energy electrons in the lobes of 
powerful radio galaxies and quasars.
This would be consistent with the 
optical emission from hot spots
via the SSC process (3C 295, Brunetti, 2001;
3C 196, Hardcastle, 2001) indicating the presence of electrons
with $\gamma$ of several hundred in the compact regions, 
so that their energy would be
degraded into the $\gamma \sim 100$ range by adiabatic losses
suffered while expanding into the radio lobes.
It also follows that the extension of the electron energy spectrum
toward lower energies should not be neglected when addressing the total 
energy content in the radio lobes (Brunetti et al., 1997).

Since the far IR to optical flux of the nucleus of 3C 207 can
be directly estimated, one can compute
the magnetic field strength in the radio lobe by the combined
radio--synchrotron and IC X--ray fluxes.
Depending on the assumed electron spectrum at lower energies, 
the magnetic field strength estimated in the C--lobe 
is within a factor of $\sim 1.1-2.2$ smaller  
than that computed under the equipartition hypothesis 
and the ratio between electron and magnetic field energy 
densities falls in the range $\sim 1-10$.
A similar result, but further on closer to equipartition, has
been obtained in the already mentioned 
case of the compact radio galaxy
3C 295 (Brunetti et al. 2001), whereas
past ROSAT HRI and ASCA IC measurements of very extended radio sources
have evidenced possible ratios between electron and magnetic 
field energy densities up to $\sim 80$ 
(e.g. Cen B, Tashiro et al., 1998, 3C 219, Brunetti et al., 1999;
Fornax A, Tashiro et al., 2001).
It is tempting to speculate that the dominance 
of the electron energy density in the
radio lobes is related to the evolution and
linear scale of the radio sources, as  
also indicated by the studies of 
synchrotron spectral ageing (e.g. Blundell \& Rawlings 2000).
Clearly more {\it Chandra} and XMM observations of strong radio 
galaxies and steep spectrum radio loud quasars
are required to test this scenario.

$\bullet$
The case of the knot is particularly interesting: it has a radio flux
similar to that of the hot spot but it is much more luminous in the
X--ray band (0.1--10 keV luminosity $\simeq 7.2 \times 10^{43}$erg
s$^{-1}$).
We have shown that, although the X--ray flux may be reproduced with an
SSC model by a strong departure from 
the equipartition condition (a magnetic
field strength a factor of $\sim 6-7$ times lower than the equipartition
value is required), the X--ray spectrum is not compatible with the
SSC scenario since it is much harder 
($\Delta \alpha \sim 0.7$) than the radio 
synchrotron spectrum.
Assuming that the X--ray and the radio fluxes are of
IC and synchrotron origin, respectively,  
the difference between the two
spectral slopes clearly indicates that  
the energy distributions of the electrons involved
in the two processes are different.

Since the knots (and the hot spots as well) are
believed to be the sites of strong particle acceleration,
we have worked out (Sect.3.1) an analytic simple model for the
re--accelerated electron spectrum injected in the post--shock region
and its time evolution. We find that the X--ray properties of the
knot are well accounted for by the IC scattering of CMB photons
(EIC model) under approximate equipartition conditions
if a relativistic
bulk motion with Lorentz factor $\Gamma_{\rm bulk} > 5$
and an angle with the line of sight $\theta_{\rm bulk} \leq 10^o$
are assumed.
With these parameters the observed X--rays are produced by
$\gamma_{\rm EIC} \sim 50$ electrons which pertain to the flatter
portion of the electron spectrum as required by the 
hard X--ray spectrum ($\gamma_{\rm low} < 100$ is required to match
the X--ray spectral shape).
This scenario is also in qualitative agreement with the
X--ray one sideness of the jet.
As 3C 207 is a steep spectrum radio quasar it is unlikely
that its radio axis forms an angle 
$\theta_{\rm i} \leq 10^o$ with the
line of sight. 
This might appear in contrast with the 
lower bound of $\theta_{\rm bulk}$.
We notice, however, that both the large scale radio structure 
and the main radio jet are relatively distorted (Fig.1).
In particular, the position of the radio knot appears to be shifted 
with respect to the line joining the innermost
radio jet and the hot spot of about $\sim 5^o$ on the plane of the sky.
Thus, it might very well be that the knot moves in a direction 
few degrees closer to the line of sight 
than the direction of the innermost radio jet.
Relativistic boosting in radio loud
objects at large distances from the nucleus has been recently 
invoked by Tavecchio et al.(2000) and Celotti et al.(2001).
These authors have independently 
interpreted the X--ray emission from the main 
knot of the flat spectrum radio loud quasar PKS 0637$-$752 as due to
IC scattering of the CMB photons and derive a relativistic
bulk motions $\Gamma_{\rm bulk} \sim 14$, which is much stronger
than that invoked in the present paper for 3C 207.
In the case of PKS 0637$-$752 
the synchrotron and X--ray spectral shapes are similar 
(Chartas et al., 2000) so that 
by adopting our procedure the 
low energy cut--off in the electron spectrum 
would result at very low energies ($\gamma_{\rm low}$\ltsim$30$).

$\bullet$
The X--ray flux of the hot spot is a factor of
$\sim$6.5 lower than that of the knot 
(0.1--10 keV luminosity $\simeq 1.1\times 10^{43}$erg s$^{-1}$),
about 40 net counts are detected and the spectrum
is not very well constrained.
It can be reproduced by either a SSC model 
or an EIC model with moderate boosting.
In both cases the derived 
rest frame magnetic field strength
is a factor of $\sim 2$ smaller than the 
equipartition value.
With the exception of the western hot spot of Pic A 
(Wilson et al.2001), 
approximate equipartition conditions (i.e. within a factor
of 2 in field) have been found in 
a number of detected X--ray hot spots   
(e.g. Cyg A, Harris et al. 1994, Wilson et al. 2000;
3C 123, Hardcastle et al. 2001a) all of which are
well fitted by SSC model.

\begin{acknowledgements}

This work is partially supported by 
the Italian Space Agency (ASI) under the contract
ASI-ARS-99-75 and by the Italian Ministry for
University and Research (MURST) under grant 
Cofin98-02-32.

\end{acknowledgements}

\appendix{}
\section{Model calculations}

\subsection{Synchro-self-Compton formulae}

Synchro-self-Compton process, i.e. inverse Compton scattering of
synchrotron radiation by the synchrotron--emitting electrons, 
is a well known radiative process in astrophysics.
The relevant astrophysical formulae published so far are generally
calculated under the approximation that both the electron energy
distribution and the synchrotron spectrum can be well described
by power laws (e.g. Jones, O'Dell, Stein 1974; Gould 1979; Marscher 1983).

In this Appendix we report semi--analytical
SSC equations used in the
model calculations and obtained 
by taking into account the correct synchrotron
spectrum and a general electron energy distribution
$N(\gamma)=K_{\rm e} f(\gamma)$.

Let be $\dot{n}_{\rm ph}^{\rm s}(r,\nu_{\rm s})$ 
the rate per unit volume of emitted synchrotron photons 
in the frequency interval 
$\nu_{\rm s}$--$\nu_{\rm s}+d\nu_{\rm s}$ per unit volume 
at a distance $r$ from the center.
Then the photon density at
distance $s$ from the center is given by:

\begin{equation}
N_{\rm ph}^{\rm s}(s,\nu_s)= {1\over{ 4 \pi c}} 
\int_{\rm Vol}
d^3 r {{ \dot{n}_{\rm ph}^{\rm s}(r,\nu_{\rm s}) }\over{
|{\bf r}-{\bf s}|^2}}
\label{ssc_start}
\end{equation}

\noindent
For a spherically homogeneous source, it is  
$\dot{n}_{\rm ph}^{\rm s}(r,\nu_{\rm s})=
\dot{n}_{\rm ph}^{\rm s}(\nu_{\rm s})$ so that:

\begin{equation}
N_{\rm ph}^{\rm s}(x,\nu_{\rm s})= 
{{\dot{n}_{\rm ph}^{\rm s}(\nu_{\rm s}) R}\over{ 2 c}} 
\{ 1+ {1\over 2} \left( {1\over x} -x \right)
{\rm ln} \left( {{1+x}\over{1-x}} \right) \}
\label{ssc_n}
\end{equation}

\noindent
$x= s/R$ being the distance in unit of the source 
radius.
 
\noindent
The inverse Compton luminosity produced by the scattering of the
synchrotron photons is obtained by convolving
the electron and seed photon energy distributions with the isotropic
Compton kernel (e.g. Blumenthal \& Gould 1970) and by integrating
over the volume of the source; one has :

\begin{eqnarray}
L^{ssc}(\nu) = 
2 \pi^2 r_0^2 c h \nu^2 R^4 
\int_0^1 x^2 dx 
\int \int {{
d\gamma d\nu_{\rm s} }\over
{ 
\nu_{\rm s}^2 \gamma^4 }}
\times \nonumber\\
N_{\rm e}(\gamma) 
N_{\rm ph}^{\rm s}(x,\nu_{\rm s})
\left[ 2 {\rm ln} \left(
{{\nu }\over{ 4 \nu_{\rm s} \gamma^2}} \right)
+1 + 4 \gamma^2 {{ \nu_{\rm s} }\over{ \nu }} 
- {{ \nu^2}\over{ 2 \nu_{\rm s} \gamma^2}} \right]
\label{ssc1}
\end{eqnarray}

\noindent
By performing the volume integral and  
Eq.(\ref{ssc_n}) one has :

\begin{eqnarray}
L^{ssc}(\nu) = {{ 3 \pi r_0^2}\over 8}
K_e R \nu^2  
\int d\nu_{\rm s} {{L^s(\nu_{\rm s}) }\over
{\nu_{\rm s}^3}}
\int_{{1\over 2}\sqrt{ {{\nu }\over{ \nu_{\rm s} }} } } 
^{\gamma_{\rm c} }
d\gamma {{f(\gamma)}\over{\gamma^4}} \nonumber\\
\times \left[ 2 {\rm ln} \left(
{{\nu }\over{ 4 \nu_{\rm s} \gamma^2}} \right)
+1 + 4 \gamma^2 {{\nu }\over{ \nu_{\rm s} }}
- {{ \nu }\over{ 2 \nu_{\rm s} \gamma^2}} \right]
\label{ssc2}
\end{eqnarray} 

\noindent
$L^s(\nu_{\rm s})$ 
being the monochromatic synchrotron luminosity and
$\gamma_{\rm c}$ the largest energy of the electrons.
From the synchrotron emissivity formula (e.g., Pacholczyk, 1970), 
the electron number density $K_{\rm e}$
can be expressed in terms of
synchrotron luminosity at a given frequency, $\tilde{\nu_{\rm s}}$, 
and magnetic field strength $B$. One obtains:

\begin{eqnarray}
L^{ssc}(\nu ) = {{3 \sqrt{3} }\over{32}}
{{ m c^2}\over{e^3}} \left({{r_0}\over{R}} \right)^2 
\nu^2 {{ {\cal K}(B) }\over{B}} 
\int d\nu_{\rm s} {{L^{\rm s}(\nu_{\rm s}) }\over
{{\nu_{\rm s}}^3}} \times \nonumber\\
\int_{{1\over 2}\sqrt{ {{\nu }\over{ \nu_{\rm s} }} } }
^{\gamma_{\rm c} }
d\gamma {{f(\gamma)}\over{\gamma^4}} 
\left[ 2 {\rm ln} \left(
{{\nu }\over{ 4 \nu_{\rm s} \gamma^2}} \right)
+1 + 4 \gamma^2 {{\nu }\over{ \nu_{\rm s} }}
- {{ \nu }\over{ 2 \nu_{\rm s} \gamma^2}} \right]
\label{ssc3}
\end{eqnarray} 

where

\begin{equation}
{\cal K}(B)=
{{ L^{\rm s}(\tilde{\nu_{\rm s}})/[4 \pi R^3 /3 ] }\over{
\int_1^{\gamma_{\rm c}} d\gamma \int_0^{\pi /2} d\theta \sin^2 \theta
f(\gamma) 
F\left[ \left( {{\gamma_{\rm c} }\over{ \gamma}} \right)^2
{{\tilde{\nu_{\rm s}} / \nu_{\rm s}^{\rm c}}\over{\sin \theta }} \right] 
}}
\label{ssc_fun}
\end{equation}

\noindent
$F[\,\,]$ the synchrotron Kernel (e.g., Pacholczyk, 1970) and  
$\nu_{\rm s}^{\rm c}({\rm MHz})
\simeq 4.2 \times 10^{-6} \gamma_{\rm c}^2
B(\mu{\rm G})$, the cut--off frequency of the
synchrotron spectrum.

\noindent
If the source is moving 
with a bulk motion Lorentz factor $\Gamma_{\rm bulk}$ about an 
angle $\theta_{\rm bulk}$ with respect to the observer, 
the received flux $F^{\rm rec}$ 
is given by (e.g., Rybicki \& Lightman 1979) :

\begin{equation}
F^{\rm rec} (\nu ) =
L \left({{ \nu }\over { { \cal{D} } }} \right) {\cal{D}}^{3}
[ 4 \pi d^2 k(z)]^{-1} 
\label{mah}
\end{equation}

\noindent
$d$ is the luminosity distance, $k(z)$ provides for
the cosmological k--correction and ${\cal D}$, the
Doppler factor :

\begin{equation}
{\cal D}=
\left\{
\Gamma_{\rm bulk} (1- \beta_{\rm bulk} 
\cos\theta_{\rm bulk} ) \right\}^{-1}
\label{d}
\end{equation}

\noindent
From Eqs.(\ref{ssc3}--\ref{d}) the received SSC flux ($F^{\rm ssc}$)
is related to the synchrotron flux ($F^{\rm s}$) by:

\begin{eqnarray}
F^{ssc}(\nu_{\rm c}) = {{3 \sqrt{3} \pi}\over{8}}
{{ m c^2}\over{e^3}} \left( {{r_0}\over{R}} \right)^2  \nu^2
{{ {\cal K}(B,{\cal D}) }\over{B}} 
\int d\nu_{\rm s} {{F^{\rm s}(\nu_{\rm s}) }\over
{{\nu_{\rm s}}^3}} \times \nonumber\\
\int_{{1\over 2}\sqrt{ {{\nu}\over{ \nu_{\rm s} }} } }
^{\gamma_{\rm c} }
d\gamma {{f(\gamma)}\over{\gamma^4}} 
\left[ 2 {\rm ln} \left(
{{\nu}\over{ 4 \nu_{\rm s} \gamma^2}} \right)
+1 + 4 \gamma^2 {{\nu}\over{ \nu_{\rm s} }}
- {{ \nu }\over{ 2 \nu_{\rm s} \gamma^2}} \right]
\label{f_ssc}
\end{eqnarray} 

\noindent
where all the frequencies are in the observer frame,  

\begin{equation}
{\cal K}(B,{\cal D})=
{{  d^2 k(z) {{\cal D}}^{-3} F^{\rm s}(\tilde{\nu_{\rm s}}) /
[4 \pi R^3 /3 ] }\over{
\int_1^{\gamma_{\rm c}} d\gamma \int_0^{\pi /2} d\theta \sin^2 \theta
f(\gamma) 
F\left[ \left( {{\gamma_{\rm c} }\over{ \gamma}} \right)^2
{{
\tilde{\nu_{\rm s}} 
/ \nu_{\rm s}^{\rm c}}\over{\sin \theta }} \right] 
}}
\label{dssc_fun}
\end{equation}

\noindent
where the synchrotron cut--off frequency is measured in the observer
frame and 

\begin{equation}
\gamma_c \simeq 488 \times 
\left[ {{ \nu_{\rm s}^{\rm c}(\rm MHz) }\over
{ B(\mu{\rm G}) {\cal D} }} \right]^{1/2}
\label{dgc}
\end{equation}

\noindent
Given an electron energy distribution $f(\gamma)$ 
(Eqs.\ref{n_fin1}--\ref{n_fin3}) and a Doppler factor
${\cal D}$, we calculate the
synchrotron spectrum as received by the observer.
Then we fit the observed synchrotron spectrum obtaining 
$\nu_{\rm s}^{\rm c}$ and, for a 
given value of $B$, we derive $\gamma_{\rm c}$.
Finally, we apply Eqs.(\ref{f_ssc}--\ref{dssc_fun}) and
calculate the synchro--self--Compton spectrum as received
by the observer and compare it with the {\it Chandra} data.

\subsection{External inverse Compton formulae}

In this Appendix we derive semi--analytic formulae
for the IC scattering between external photons and an
electron population in a homogenous sphere moving
with a Doppler factor ${\cal D}=1/[ \Gamma_{\rm bulk}
(1- \beta_{\rm bulk} \cos \theta_{\rm bulk} ]$.
Dermer (1995) first derived 
approximate equations holding in the ultra--relativistic
case (i.e., $\gamma >> 1$), 
by assuming that the seed photons in the blob frame
are coming from a direction opposite to the velocity of the blob, 
and assuming a power law energy distribution of the scattering
electrons.

\begin{figure}
\includegraphics{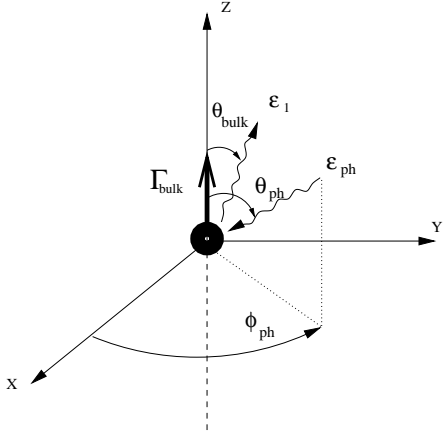}
\vspace{7 cm}
\caption[]{The assumed geometry is reported with the relevant quantities.
The blob moves along the
z--axis, whereas the scattered photons ($\epsilon_1$) 
move in the y--z plane.} 
\end{figure}

Here we work out a new set of equations 
three main reasons:

\noindent
-- in general, the electron spectrum can be more complicated than 
a simple power law (e.g., Eqs.\ref{n_fin1}--\ref{n_fin3})
and the above equations cannot be used; 

\noindent
-- in the case of moderate boosting or de--boosting, 
the direction of the seed photons in the blob frame 
cannot be simply approximated with the 
velocity of the blob and the correct 
photon angular distribution should be taken into account;

\noindent
-- in the case $\Gamma_{\rm bulk} >>1$ the X--rays from
the IC scattering of external photons
of frequency $>> 10^{11}$Hz are mainly contributed by mildly
relativistic electrons 
and the assumption $\gamma >> 1$ should be released;

We assume a scattering geometry as in Fig.A1; in the following
the primed quantities are referred to the blob frame. 
In this case the four--vectors in the observer frame
of the incoming and scattered
photons are given by:

\begin{eqnarray}
P_{\epsilon_{\rm ph}}=
{{ \epsilon_{\rm ph}}\over{c}}
(1,-\sin\theta_{\rm ph} \cos\phi_{\rm ph},
-\sin\theta_{\rm ph} \sin\phi_{\rm ph}, \nonumber\\
-\cos\theta_{\rm ph})
\end{eqnarray}

\begin{equation}
P_{\epsilon_1}=
{{ \epsilon_1}\over{c}}
(1,0,
\sin\theta_{\rm bulk},
\cos\theta_{\rm bulk})
\end{equation}

\noindent
The number density of the incoming photons in the
blob frame is given by: 

\begin{equation}
{{ d n^{\prime} }\over{d\epsilon_{\rm ph}^{\prime}
d\Omega_{\rm ph}^{\prime} }}=
{{ d n }\over{d\epsilon_{\rm ph}
d\Omega_{\rm ph} }}
\left[ \Gamma (1+ \beta_{\rm bulk} \cos\theta_{\rm ph} ) \right]^2
\label{n}
\end{equation}

\noindent
If the incoming external photons are those of the CMB, one has:

\begin{equation}
{{ d n^{\prime} }\over{d\epsilon_{\rm ph}^{\prime}
d\Omega_{\rm ph}^{\prime} }}=
{ { 2 {\epsilon^{\prime}_{\rm ph}}^2 }\over{({\rm hc})^3 }}
\left[
{\rm exp}\{ {{ \Gamma \epsilon^{\prime}_{\rm ph} 
(1 - \beta_{\rm bulk} \cos\theta^{\prime}_{\rm ph})}\over{ 
{\rm kT} }} \} -1 \right]^{-1}
\label{nf}
\end{equation}

\noindent
where $T$ is the local CMB temperature and ${\rm k}$ is
the Boltzmann constant.
The flux received by the observer is given by Eq.(A.7) :

\begin{equation}
F^{\rm rec}(\epsilon_1,\theta_{\rm bulk})=
{{
L^{\prime}(\epsilon_1^{\prime} \, , \, \Omega_{\rm i}^{\prime})
}\over{
k(z) d^2 
}}
{{ \cal D }}^3
\end{equation}

\noindent
where $L(\epsilon_1^{\prime} \, , \, \Omega_{\rm i}^{\prime})
=4 \pi R^3 /3 \times j^{\prime}(\epsilon_1^{\prime} \, , \, 
\Omega_{\rm i}^{\prime})$ ($j^{\prime}$ the emissivity per unit
solid angle) is the IC luminosity per unit solid angle and
energy emitted in the direction of the observer.

The IC emissivity is obtained by integrating the IC cross section
(Thomson approximation) given the energy and angular distribution of the 
incoming photons and the relativistic electron spectrum.
For the details of the calculation of the general IC emissivity
in the case of anisotropic angular distribution of the
incoming photons, such as is the present case due to the bulk motion, 
we remind the reader to Brunetti (2000).
The IC emissivity per unit solid angle in the direction
$\theta^{\prime}_{\rm bulk}$, at an energy $\epsilon^{\prime}_1$, 
is given by:

\begin{eqnarray}
j^{\prime}(\epsilon_1^{\prime},\theta_{\rm bulk}^{\prime})
= {{ K_e r_o^2 c}\over {4}} \int d\Omega_{\rm ph}^{\prime}
d\epsilon_{\rm ph}^{\prime}
{{ d n^{\prime} }\over{d\epsilon_{\rm ph}^{\prime}
d\Omega_{\rm ph}^{\prime} }}  \nonumber\\
\times \left( {{\epsilon_1^{\prime} }\over
{\epsilon_{\rm ph}^{\prime} }} \right)^2
\{ 2 {\cal I}_0 
\left[ \left( {{\epsilon_1^{\prime} }\over{
\epsilon_{\rm ph}^{\prime} }} \right)^2
- {{ 2 {\cal L} \epsilon_1^{\prime} }\over{\epsilon_{\rm ph}^{\prime} }}
+1 \right]^{-{{1\over2}} } \nonumber\\
- 2 [1-{\cal L}]^2 \left( 1+ {{\epsilon_{\rm ph}^{\prime} }\over{
\epsilon_1^{\prime} }} \right) {\cal I}_{3 \over 2} + 
{[1-{\cal L}]}^3 \times \nonumber\\
\times \left[ \left( 1+ {{\epsilon_{\rm ph}^{\prime} }\over{
\epsilon_1^{\prime} }} \right) ( 3 {\cal L} -{3\over 2} )
- {3 \over 2} \left( ( {{\epsilon_{\rm ph}^{\prime} }\over{
\epsilon_1^{\prime} }} )^2 +
{{ \epsilon_1^{\prime} }\over{ \epsilon_{\rm ph}^{\prime} }}
\right) \right] {\cal I}_{5\over 2} \nonumber\\
+{5\over 2} [1-{\cal L}]^5 
\left[ 3 \left( 1+ {{\epsilon_{\rm ph}^{\prime} 
}\over{\epsilon_1^{\prime} }} 
\right) + ( {{\epsilon_{\rm ph}^{\prime} }\over{
\epsilon_1^{\prime} }} )^2+
{{ \epsilon_1^{\prime} }\over{ \epsilon_{\rm ph}^{\prime} }} \right]
{\cal I}_{7\over 2} \}
\label{jprime}
\end{eqnarray}

\noindent
where, with the adopted geometry of Fig.A1, one has :

\begin{equation}
{\cal L}= -\cos(\theta^{\prime}_{\rm bulk}) \cos(\theta^{\prime}_{\rm ph})
- \sin(\theta^{\prime}_{\rm bulk}) \sin(\theta^{\prime}_{\rm ph})
\sin(\phi^{\prime}_{\rm ph})
\label{l}
\end{equation}

\noindent
whereas, 
the functions ${\cal I}$ are :

\begin{equation}
{\cal I}_0=
\int_{\gamma_{\rm min}}
{{ \gamma^{-2} f(\gamma) d\gamma}\over{
(1- \gamma^{-2} )^{1/2} }}
\end{equation}

\begin{equation}
{\cal I}_{3/2}=
\int_{\gamma_{\rm min}}
{{ \gamma^{-1} f(\gamma) d\gamma}\over{
(1- \gamma^{-2} )^{1/2} 
[ \gamma^2 (1- {\cal L})^2 + 1 - {{\cal L}}^2 ]^{3/2}
}}
\end{equation}

\begin{equation}
{\cal I}_{5/2}=
\int_{\gamma_{\rm min}}
{{ \gamma^{-1} f(\gamma) d\gamma}\over{
(1- \gamma^{-2} )^{1/2} 
[ \gamma^2 (1- {\cal L})^2 + 1 - {{\cal L}}^2 ]^{5/2}
}}
\end{equation}

\begin{equation}
{\cal I}_{7/2}=
\int_{\gamma_{\rm min}}
{{ \gamma f(\gamma) d\gamma}\over{
(1- \gamma^{-2} )^{1/2} 
[ \gamma^2 (1- {\cal L})^2 + 1 - {{\cal L}}^2 ]^{7/2}
}}
\end{equation}

\noindent 
and the minimum energy of the electrons involved in the scattering
is given by:

\begin{equation}
\gamma_{\rm min}=
\left\{
1+ {{ {\cal D} ( \epsilon_1 / {\cal D} -\epsilon^{\prime}_{\rm ph} )^2}\over
{ 2 \epsilon^{\prime}_{\rm ph} \epsilon_1 (1- {\cal L})  }}
\right\}^{1\over 2}
\end{equation}

\noindent
so that, expressed in observer frame, the received
IC power from a unit volume (per unit energy 
and solid angle) is:

\begin{eqnarray}
F^{\rm rec}(\omega_1, \theta_{\rm bulk})= 
{2\over 3} {{ \pi r_o^2 K_e}\over{ {\rm h}^3 {\rm c}^2  }}
{{ R^3 }\over{ d^2}}
({\rm kT} )^3 {\cal D} \omega_1^2 
\int {{ d\Omega^{\prime}_{\rm ph} }\over{
D(\theta^{\prime}_{\rm ph}) }} \times 
\nonumber\\
\int {{ dx }\over{ e^x -1 }} \{ 2 {\cal I}_0 
\left[ y_1^2(x,\theta^{\prime}_{\rm ph})
-2 {\cal L} y_1(x,\theta^{\prime}_{\rm ph})
+1 \right]^{- {{ 1\over2}} }
\nonumber\\
-2 [1-{\cal L}]^2 
\left( 1+ {1\over{ y_1(x,\theta^{\prime}_{\rm ph}) }}
\right) {\cal I}_{3 \over 2}  \nonumber\\
+ {[1-{\cal L}]}^3 \times 
[ \left( 1+ 
{1\over{ y_1(x,\theta^{\prime}_{\rm ph}) }}
\right) ( 3 {\cal L} -{3\over 2} )-
\nonumber\\
{3 \over 2} \left( {1\over{ y_1^2(x,\theta^{\prime}_{\rm ph}) }}
+y_1(x,\theta^{\prime}_{\rm ph}) 
\right)  ] {\cal I}_{5\over 2}
\nonumber\\
+{5\over 2} [1-{\cal L}]^5 
[ 3 \left( 1+ 
{1\over{ y_1(x,\theta^{\prime}_{\rm ph}) }}
\right) +
\nonumber\\
{1\over{ y_1^2(x,\theta^{\prime}_{\rm ph}) }}
+y_1(x,\theta^{\prime}_{\rm ph})
]
{\cal I}_{7\over 2} \}
\end{eqnarray}

\noindent
where $\omega_1 = \epsilon_1/{\rm kT}$,
$y_1(x,\theta^{\prime}_{\rm ph})= \epsilon^{\prime}_1 / 
\epsilon^{\prime}_{\rm ph} =
{{\omega_1}\over{ x }} 
{{ D(\theta^{\prime}_{\rm ph}) }\over{ {\cal D} }}$
and
$D(\theta^{\prime}_{\rm ph})=
\Gamma (1 - \beta_{\rm bulk} \cos\theta^{\prime}_{\rm ph})$.

\noindent
In the ultrarelativistic case, 
$y_1 = \epsilon^{\prime}_1 / \epsilon^{\prime}_{\rm ph} >> 1$,
one has:

\begin{eqnarray}
F^{\rm rec}(\omega_1, \theta_{\rm bulk}) \rightarrow
{2\over 3} {{ \pi r_o^2 K_e}\over{ {\rm h}^3 {\rm c}^2  }}
{{ R^3 }\over{ d^2}}
({\rm kT} )^3 {\cal D} \omega_1^2 
\int {{ d\Omega^{\prime}_{\rm ph}  }\over{
D(\theta^{\prime}_{\rm ph}) }} \times\nonumber\\
\int dx
 { { dx }\over{ e^x -1}} \times 
\{
{ 2 \over { y_1(x,\theta^{\prime}_{\rm ph}) }} 
\int \gamma^{-2} f(\gamma)d\gamma
\nonumber\\
-{ 2\over{1- {\cal L} }} \int \gamma^{-4} f(\gamma)d\gamma 
\nonumber\\ +
y_1(x,\theta^{\prime}_{\rm ph})
\left( 1- {\cal L} \right)^{-2} \int \gamma^{-6} 
f(\gamma)d\gamma \}
\label{ur}
\end{eqnarray}

\noindent
This formula is of immediate application once the electron
number density ($K_e$) is constrained by the
synchrotron spectrum (Eq.\ref{dssc_fun}).

\begin{figure}
\includegraphics{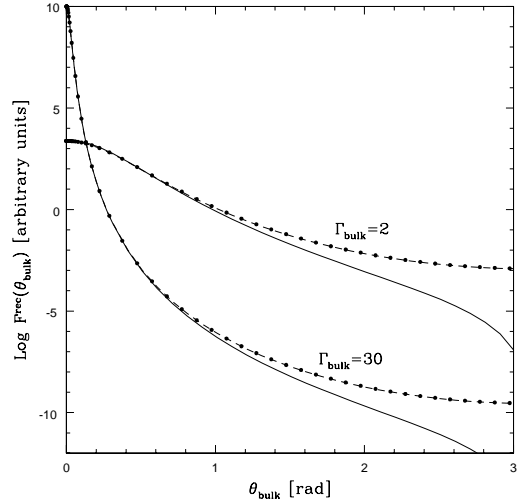}
\vspace{7 cm}
\caption[]{The angular distribution of the received monochromatic power 
as predicted in the ultra--relativistic approximation for a power low
energy distribution of the scattering electrons 
(Eq.A.26; points) is compared
with a ${ {\cal D}}^{4+2\alpha}$ law (dashed lines), and
with the Dermer (1995) formula (solid lines).} 
\end{figure}

\noindent
In the simple case 
$f(\gamma)=\gamma^{-\delta}$, one has:

\begin{eqnarray}
F^{\rm rec}(\omega_1, \theta_{\rm bulk}) \rightarrow
{{ \pi r_o^2 K_e  R^3 }\over{3 {\rm h}^3 {\rm c}^2 d^2 }}
({\rm kT})^{ 3 }
{{ 2^{ \alpha+3 }
[\delta^2 + 4\delta +11]}\over{(\delta+1)(\delta+3)
(\delta+5) }} \nonumber\\
\times \omega_1^{-\alpha} { {\cal D} }^{3+\alpha}
I  \int d\Omega^{\prime}_{\rm ph}
{{ [1- {\cal L} ]^{\alpha+1} }\over
{ { {\cal D}_{\rm ph} }^{\alpha+3} }}
\label{urpl}
\end{eqnarray}

\noindent
where $\alpha = (\delta-1)/2$, and
 
\begin{equation}
I= \int_0^{\infty}
dx {{ x^{\alpha+2} }\over
{ e^{x} -1 }}
\end{equation}

\noindent
We notice that 
Eq.(\ref{urpl}) is slightly different from the Dermer (1995)
result (Fig.A2), whereas it coincides for all $\theta_{\rm 
bulk}$ with the ultra--relativistic analytical result 
(for a power law electron spectrum), 
$F^{\rm rec} \propto \omega_1^{-\alpha}
{ {\cal D}}^{4+2\alpha}$, obtained without assuming that the seed photons
in the blob frame are coming in the direction opposite to the blob
velocity (as published by Georgantopulos et al.2001 while this paper
was in preparation).

\end{document}